\documentclass[aps,prc,twocolumn,superscriptaddress,showpacs,floatfix]{revtex4}
\usepackage[dvips]{graphicx}
\usepackage{amsmath,amssymb}

\begin{document}
\title{ Microscopic calculation of the wobbling excitations
       by using the Woods-Saxon potential as a nuclear mean-field }

\author{Takuya Shoji}
\affiliation{Department of Physics, Graduate School of Sciences,
Kyushu University, Fukuoka 812-8581, Japan}

\author{Yoshifumi R. Shimizu}
\affiliation{Department of Physics, Graduate School of Sciences,
Kyushu University, Fukuoka 812-8581, Japan}

\date{\today}

\begin{abstract}

The wobbling excitations of the triaxial superdeformed (TSD) bands
in the Lu and Hf region are studied
by the microscopic framework of the cranked mean-field and
the random-phase approximation (RPA).
In contrast to the previous works, where the Nilsson potential
was used, the more realistic Woods-Saxon potential is employed
as a nuclear mean-field.
The wobbling-like RPA solutions have been found systematically
in the nuclei studied and their characteristic properties
are investigated in details.
This confirms the wobbling phonon excitations in TSD nuclei
from the microscopic calculations.
The result of $B(E2)$ values indicates that the triaxial deformation
is increasing as a function of spin in the observed TSD bands in $^{163}$Lu.

\end{abstract}

\maketitle

\section{Introduction}
\label{sect:intro}

The observation of the wobbling excitations~\cite{wob163Lu1,wob163LuTwo}
(see also Ref.~\cite{Hamamoto01,Hageman04})
renewed the interest in the study of nuclear rotational motions.
The nuclear wobbling motion~\cite{BMtext75} is a quantized motion
of the triaxial rotor and appears as a multiple-band structure,
in which consecutive rotational bands are connected by strong
$E2$ transitions with each other.
Until now, the multiple rotational band structure
characteristic to the wobbling phonon excitation has been observed
at the high-spin excited states in some Lu
isotopes~\cite{wob163Lu1,wob163Lu2,Hamamoto01,wob163Lu3,wob163LuTwo,%
wob165Lu,wob167Lu,wob161Lu} around $^{163}$Lu.
Nuclei in the Lu and Hf region, including these Lu isotopes,
have been predicted to be strongly deformed
with pronounced triaxiality~\cite{Rag89,Ab90,BenRyd04,Bengt}
at high-spin states, and the associated rotational bands
are called triaxial superdeformed (TSD) bands; i.e., the wobbling
structure is composed of these TSD bands.
In fact the lifetime measurements in these nuclei~\cite{wobLuQ,wob163LuQ}
revealed that the rotational $E2$ transition probabilities
inside the bands are typically about 500 Weisskopf units,
and those between the bands associated with the wobbling phonon excitations
are about 100 Weisskopf units.
These are one of the largest out-of-band $B(E2)$ and
believed to be the evidence of the nuclear wobbling motion.
Although several candidates of the TSD bands have been observed in
even-even Hf isotopes~\cite{TSD168Hf,TSD170Hf1,TSD170Hf2,TSD172Hf,TSD174Hf},
there is no definite evidence of the wobbling excitation yet.

The nuclear wobbling motion was first predicted by using
the simple macroscopic rotor model~\cite{BMtext75}.
The experimental data observed in Lu isotopes are investigated
by the particle-rotor model~\cite{Ham02,HH03,TST06,TST08},
because a $i_{13/2}$ quasiproton exists in the proton-odd Lu nuclei.
It is, however, noted that the properties of the observed out-of-band
$E2$ transitions suggest that the triaxial deformation is of
the so-called positive-$\gamma$ shape in the Lund convention~\cite{NR95},
the sign of which is opposite to that of Ref.~\cite{BMtext75};
namely the moment of inertia about the shortest axis of triaxial
deformation is the largest.  This conflicts with the irrotational moments
of inertia which are natural for the macroscopic rotor model,
where the moment of inertia is largest about the intermediate axis.
In principle, the triaxiality for the quadrupole shape and that for
the moments of inertia can be taken differently in the macroscopic model;
the so-called ``$\gamma$-reversed'' moments of inertia,
i.e., those about the shortest and intermediate axes are interchanged,
are used in Refs.~\cite{Ham02,HH03}, while the rigid-body moments of
inertia that are consistent with the positive $\gamma$ shape
are recommended in Refs.~\cite{TST06,TST08}.
Since the moments of inertia are the basic parameters (inputs) of
the macroscopic model, microscopic approaches are necessary
to investigate this problem.

From the theoretical point of view, how the rotor model emerges
out of the collective rotational motions of constituent nucleons is
an interesting and long-standing problem (see e.g. Ref.~\cite{RingSchuck}).
Most of nuclei are axially-deformed in their ground states
and the collective rotation is of one-dimensional nature
(rotation about only one axis).
In contrast, the wobbling excitations correspond to tilting
the main rotation axis from the principal axis of deformation,
and the rotation is of three-dimensional nature.
Therefore, the microscopic study of the wobbling motion,
which is characteristic in the rotor model, is expected to give
a new insight to the problem of how the individual nucleons
form and affect the triaxial nuclear rotor.
The key to relate the motions of nucleons to the collective
rotation is the introduction of the ``body-fixed'' or
the principal axis (PA) frame~\cite{VC70},
in which no collective rotations exist.
The condition of the body-fixed frame defines
the constraints to the many-body system,
and combined with the redundant rotor degrees of freedom
the original many-body problem is recovered.
The idea of this ``collective coordinate method'' in quantum many-body
systems was put forward in Ref.~\cite{BesKur90}, especially
for the nuclear triaxial rotor problem in~\cite{KBC90,GB94}.
The rigorous quantum mechanical treatment of the constraints is
a difficult problem, but Marshalek showed~\cite{Mar79}
by taking the small amplitude limit of the time-dependent HFB
(TDHFB) theory that
the transformation to the PA frame is straightforward within
the order of the random-phase approximation (RPA),
see also Refs.~\cite{JM79,Zel80}.
The semiclassical treatment of the collective rotational motion
based on the TDHFB theory
was developed in Refs.~\cite{KO81,Onishi86}, in which
the PA frame constraints are taken into account properly.
The attempts to go beyond the RPA order have been done~\cite{Kaneko92,Kaneko94}
based on the selfconsistent-collective coordinate
method~\cite{SCC-orig,SCC-YK}.

In Marshalek's theory~\cite{Mar79},
it has been shown that the RPA eigenenergy
can be written in the same form as the wobbling phonon energy
in terms of the three moments of inertia,
which can be derived either in the classical asymmetric top~\cite{LanMec} or
in its quantum analogy, the triaxial rotor model~\cite{BMtext75},
by using the small amplitude assumption consistent to the RPA.
It has been also shown~\cite{SM95} that the $E2$ transition
probabilities associated with the wobbling excitation can be written
in the same way as in the rotor model~\cite{BMtext75}
if the RPA eigenmode has a certain property.
It should, however, be stressed that the actual calculations
are performed in the cranked mean-field plus RPA
approach~\cite{Mar77,EMR80,SM83}, which is formulated
in the uniformly rotating (UR) frame
according to the usual cranking prescription.
It is the introduction of the PA frame and the transformation
from the UR to PA frame that makes it possible to interpret
a microscopic RPA eigenmode as the wobbling phonon and to derive
the three moments of inertia of the rotor microscopically.
In the theory of Ref.~\cite{Mar79} as well as those in~\cite{KO81,Kaneko94},
the PA frame is defined such that the non-diagonal part of
the (mass) quadrupole tensor should vanish identically~\cite{VC70}.
It is discussed in Refs.~\cite{BesKur90,GB94} that the PA frame condition
is a gauge fixing condition and physical observables should not depend
on its choice; for example, the three moments of inertia in the PA
(body-fixed) frame themselves are gauge-dependent and
are not the physical observables.
However, in order to obtain a clear physical picture and a relation to
the macroscopic rotor model, the introduction of the specific
PA frame condition corresponding to the nuclear rotor
is important and useful.

According to the Marshalek's wobbling theory,
microscopic RPA calculations have been performed for nuclei
in the Hf and Lu region~\cite{MSMwob1,MSMwob2,MSMwob3},
where it was found that
the RPA eigenmode that can be interpreted as the wobbling phonon
exists in each nucleus stably in a reasonable range of
mean-field deformations (see also Ref.~\cite{AND06}).
Although it is not confirmed in experiment,
the possible wobbling excitations in normal deformed nuclei
with the negative-$\gamma$ shapes have been also studied
previously~\cite{Mat90,SM95} and more recently~\cite{KN07}.
The relation between the instability of the wobbling RPA mode
and the appearance of the tilted axis cranking~\cite{FraTAC}
mean-field solution is also discussed~\cite{MO04,Mat08,AND06}.
In the previous calculations~\cite{MSMwob1,MSMwob2,MSMwob3},
the calculated out-of-band to in-band $B(E2)$ ratio was systematically
smaller by about factor two than the experimental data,
but it was shown in Ref.~\cite{SSMtri} that this is not a problem:
The triaxiality parameter in the Nilsson potential,
which was determined by the Nilsson Strutinsky calculation
and used in the RPA calculation, is too small compared
with the value estimated by the macroscopic rotor model.
If the same value as that used in the particle-rotor model analysis
is employed, the RPA calculation gives
correct magnitudes of the $B(E2)$ ratio.

In all the previous realistic calculations of the RPA wobbling
excitations mentioned above,
the Nilsson (modified oscillator) potential is used as a nuclear mean-field.
The problem of the Nilsson potential applied to high-spin cranking
calculations is well-known:  The moment of inertia is overestimated
due to the $\mbox{\boldmath$l$}^2$ potential, see e.g.~\cite{ALL76}.
The simplest way to avoid this problem is to apply
the Strutinsky renormalization to the angular momentum expectation value,
which cannot be used in the RPA calculation.
In fact the absolute value of the moment of inertia is about
30\% overestimated in the previous calculations~\cite{MSMwob1,MSMwob2,MSMwob3}.
More fundamental remedy is to include a new field-coupling in order to
restore the local Galilean invariance broken by the velocity dependent
potential like the $\mbox{\boldmath$l$}^2$ term, which was suggested
in~\cite{BMtext75} and formulated in Ref.~\cite{Kinouchi}.
Based on this formalism the correction to the cranking term was
included in the high-spin cranking calculation in~\cite{NMMS96}.
Recently, this method has been successfully employed in Ref.~\cite{KN06},
and the author of Ref.~\cite{KN06} claimed that the problem of the Nilsson
potential is solved.
However, when this method is applied, we found that no RPA solutions
corresponding to the wobbling phonon were obtained for the TSD nuclei
in the Lu and Hf region.
The reason is clear: The method is only applied to one component
along the cranking axis among three possible rotational axes.
The field-coupling to restore
the local Galilean invariance is realized as
a residual two-body interaction, which is composed of
three components corresponding to the infinitesimal rotations
about each Cartesian axis, in the original formulation~\cite{Kinouchi}.
The method to include the correction to the cranking model is
derived by using the mean-field (HF or HFB) approximation
to this residual interaction,
so that only the component of the cranking ($x$-) axis is effective.
All the three components, especially the perpendicular ($y,z$-) components,
are necessary to be included for the description of the wobbling excitation,
which is of generic three dimensional rotation.
Thus, the additional residual interaction should be included
in the RPA wobbling calculation.

The present work is along the same line as those works based
on the formalism of the cranked mean-field and the RPA,
but we use the Woods-Saxon potential
as a nuclear mean-field to avoid the problem mentioned above
because there is no velocity dependent part of the potential
(except for the spin-orbit term).
Apart from the problem of the velocity dependence,
the Woods-Saxon potential is believed to be more realistic
than the Nilsson potential.  It is desirable to confirm the existence
of the wobbling-like RPA solutions also in this mean-field potential.
This is the main subject of the present work.
The paper is organized as follows:
The basic formulation of the wobbling RPA theory~\cite{Mar79,SM95}
is reviewed in Sec.~\ref{sect:formul}.  The theory needs to be slightly
modified for the general deformed mean-field like the Woods-Saxon
potential.  The necessary modification is explained in this section.
The results of numerical calculations are presented in Sec.~\ref{sect:result},
where the detailed and systematic analyses are performed.
Section \ref{sect:sum} is devoted to the summary.
A part of the present work was presented
in a conference report~\cite{SS06}.

\section{RPA wobbling theory}
\label{sect:formul}

Although it may be desirable to perform the full selfconsistent
cranked HFB plus RPA calculation
by using a realistic two-body effective interaction,
such as one of Skyrme forces or Gogny forces,
it is still not very easy because all the symmetries except the parity
and signature are broken in the triaxial superdeformed (TSD) rotational band.
One way is to use a simple mean-field and a schematic interaction
such as the pairing plus quadrupole interaction~\cite{RingSchuck}.
We take a different approach:  Starting from a reliable
nuclear mean-field for triaxially deformed nuclei,
we construct the residual interaction that is suitable to describe
the nuclear wobbling motion.
In this way, we can investigate the properties of the wobbling mode
for any given mean-fields, e.g., its dependence on the triaxiality or
the pairing gap.
In this section, after explaining the Woods-Saxon potential
and how to construct the residual interaction,
we discuss the RPA wobbling formalism, which needs to be slightly
modified from the original one by Marshalek~\cite{Mar79}.

\subsection{Triaxially deformed Woods-Saxon potential}
\label{sect:WSpot}

The triaxially deformed Woods-Saxon potential considered in this work
is a standard one widely used in the study of
high-spin states~\cite{DMS79,RS81,NR81}:
\begin{equation}
 V_{\rm WS}(\mbox{\boldmath$r$})=V_{\rm c}(\mbox{\boldmath$r$})
 + \lambda_{\rm so} \left(\frac{\hbar}{2{M_{\rm red}}c}\right)^2
 \bigl(\mbox{\boldmath$\nabla$}V_{\rm c}(\mbox{\boldmath$r$})\bigr)
 \cdot \bigl(\mbox{\boldmath$\sigma$}
 \times \frac{1}{i}\mbox{\boldmath$\nabla$}\bigr).
\label{eq:WSpot}
\end{equation}
The first term is the central part and the second is
the spin-orbit term.
${M_{\rm red}=\frac{A-1}{A}M}$
with $A$ being the mass number and $M$ being the nucleon mass, and
$\mbox{\boldmath$\sigma$}$ is the Pauli matrices
(the nucleon spin
${ \mbox{\boldmath$s$}=\frac{\hbar}{2}\mbox{\boldmath$\sigma$}}$).
The explicit form of the central potential is
\begin{equation}
{\displaystyle
 V_{\rm c}(\mbox{\boldmath$r$})=
 \frac{V}{1+\exp(\mbox{dist}_\Sigma(\mbox{\boldmath$r$})/a)}
 },
\label{eq:WSdef}
\end{equation}
where the strength $V$ is given by
\begin{equation}
 V=-V_0\times \left(1 \pm \kappa\frac{N-Z}{A} \right),
 \quad
 \left\{\begin{array}{l} + \mbox{proton}\\ - \mbox{neutron} \end{array}
 \right.,
\label{eq:WSstreng}
\end{equation}
with $Z$ and $N$ being the proton and neutron numbers, respectively,
and
$\mbox{dist}_\Sigma(\mbox{\boldmath$r$})$ is the distance
between a given point $\mbox{\boldmath$r$}$
and the nuclear surface $\Sigma$,
with a minus sign if $\mbox{\boldmath$r$}$ is inside $\Sigma$.
The nuclear surface $\Sigma$ in this work is parametrized
by three deformation parameters, $(\beta_2,\gamma,\beta_4)$,
and is defined by the usual radius to solid-angle relation,
$r=R(\Omega)$;
\begin{equation}
\begin{array}{l}
{\displaystyle
 R(\Omega)=R_0\,c_{\rm v}(\beta_2,\gamma,\beta_4)
 }\vspace{1mm} \\
{\displaystyle
\times
\Bigl(1 +\sum_{K=0,\pm 2}a_{2K}Y_{2K}(\Omega)
 +\sum_{K=0,\pm 2, \pm 4}a_{4K}Y_{4K}(\Omega)
 \Bigr),
 }
\end{array}
\label{eq:defRS}
\end{equation}
where $c_{\rm v}(\beta_2,\gamma,\beta_4)$ is determined
by the volume conserving condition,
and the coefficients $a$'s are given by
\begin{equation}
\left\{ \begin{array}{l}
 a_{20}=\beta_2\cos\gamma,\\
 a_{22}=a_{2-2}=-\frac{1}{\sqrt{2}}\,\beta_2\sin\gamma,\\
 a_{40}=\frac{1}{6}\,\beta_4
  (5\cos^2\gamma+1),\\
 a_{42}=a_{4-2}=-\sqrt{\frac{5}{6}}\,\beta_4
  \cos\gamma\sin\gamma,\\
 a_{44}=a_{4-4}=\sqrt{\frac{35}{72}}\,\beta_4
  \sin^2\gamma.
\end{array} \right. \vspace*{2mm}
\label{eq:WSbeta}
\end{equation}
The Coulomb potential,
which is calculated by assuming the uniform charge distribution
inside the surface $\Sigma$ given by the central potential,
\begin{equation}
 V_{\rm Coul}(\mbox{\boldmath$r$})
 =\frac{3(Z-1)e^2}{8\pi R_0^3} \int\!\!\!\int_\Sigma
 \frac{(\mbox{\boldmath$r$}-\mbox{\boldmath$r$}')\cdot d\mbox{\boldmath$S$}'}
  {|\mbox{\boldmath$r$}-\mbox{\boldmath$r$}'|},
\label{eq:Coulpot}
\end{equation}
is added to the proton potential.

The potential is completely specified by the parameters,
$V_0$, $\kappa$, $R_0$, $a$, and $\lambda$.
In this work we use a set of those parameters provided
by Ramon Wyss~\cite{RWyssPrv},
among which $\kappa$ and $R_0$ are different in the central and spin-orbit
potentials.  These parameters are determined by the requirement that
the moments of inertia and the quadrupole moments
can be nicely reproduced systematically for medium and heavy nuclei
throughout the nuclear chart.
They are given in Table~\ref{tab:WSparam}.

\begin{table*}[hbtp]
\caption{ The parameters of the Woods-Saxon potential used
in this work~\cite{RWyssPrv}. Other physical constants used are
$e^2/\hbar c$=137.03602,
$\hbar c=197.32891$ MeV$\cdot$fm,
and $Mc^2=938.9059$ MeV.
$A$ in the table denotes the mass number of nucleus.
}
\label{tab:WSparam}
\begin{ruledtabular}
\begin{tabular}{cccccccc}
$V_0$ (MeV) & $\kappa_{\rm c}$ & $\kappa_{\rm so}$
 & $R_{\rm 0c}$ (fm) & $R_{\rm 0so}$ &
 & $a$ (fm) & $\lambda_{\rm so}$ \\
 \hline
53.7 & 0.63 & 0.25461
 & $1.193\,A^{1/3}+0.25$ & $0.969\times R_{\rm 0c}$ &
 & 0.68 & 26.847
\end{tabular}
\end{ruledtabular}
\end{table*}

It should be mentioned~\cite{SSMtri} that the triaxiality parameter $\gamma$
in Eq.~(\ref{eq:WSbeta}),
for which we denote $\gamma(\mbox{pot:WS})$ if necessary,
is considerably different from the one defined in terms of
the quadrupole moments, $\gamma(\mbox{den})$.
For example, $\gamma(\mbox{den}) = 20^\circ$ corresponds to
$\gamma(\mbox{pot:WS}) \approx 30^\circ$ for the case of
large deformation such as the TSD bands,
which gives important consequences on the interpretation
of the $B(E2)$ observed in the wobbling TSD bands (see Ref.~\cite{SSMtri}).
This point will be discussed in more details in Sec.~\ref{sect:BE2triaxial}.

\subsection{Residual interaction and RPA in the uniformly rotating frame}
\label{sect:ResIntUR}

The method to construct the residual interaction is discussed
in relation to the wobbling motion in Ref.~\cite{SMMprec}.
The idea is based on the decoupling of the rotational Nambu-Goldstone
(NG) modes, i.e., the angular momentum operators $J_i$ $(i=x,y,z)$,
within the RPA~\cite{PS77}. The same idea is formulated
in the context of the particle-vibration coupling theory~\cite{BMtext75}.
Although the following discussions are essentially the same as those
of Refs.~\cite{Mar79,JM79}, we recapitulate them because some of
their results are used in the later discussions in order to make clear
the point that should be modified from
the original RPA wobbling theory~\cite{Mar79}.
It should be noted that the signature, the $\pi$ rotation about
the cranking axis ($x$-axis), is a good quantum number,
and only the part of the RPA equation which transfer the signature quantum
number $\alpha$ by ${\mit\Delta}\alpha=1$, the so-called
signature $(-)$-part, is relevant to the wobbling motion.
Therefore, we consider only the corresponding part of
the residual interaction.

Starting from the general mean-field hamiltonian, $h$,
which, in the present work, is composed of
the kinetic energy, the Woods-Saxon potential, the Coulomb potential,
and the pairing part (see Sec.~\ref{sect:calMF}),
the relevant part of the residual interaction is given by
\begin{equation}
 H_{\rm res}=-\frac{1}{2}\chi_y F_y^2 - \frac{1}{2}\chi_z (iF_z)^2,
\label{eq:resInt}
\end{equation}
where the operators $F_y$ and $F_z$ are defined by
\begin{equation}
  F_y\equiv [h,iJ_y], \quad iF_z\equiv i[h,J_z].
\label{eq:defF}
\end{equation}
The strengths of the interaction are determined by
the decoupling condition of the NG modes, $J_y$ and $J_z$,
for the total hamiltonian, $H\equiv h+H_{\rm res}$,
\begin{equation}
 [H,J_{y,z}]_{\rm RPA}= [h+H_{\rm res},J_{y,z}]_{\rm RPA}=0,
\label{eq:NGdecouple}
\end{equation}
where the subscript RPA means that the commutator is evaluated within
the RPA order, leading to
\begin{equation}
 \left\{\begin{array}{l}
  1/\chi_y = \bigl\langle [[h,iJ_y],iJ_y] \bigr\rangle,
 \vspace*{1mm}\\
  1/\chi_z =- \bigl\langle [[h,J_z],J_z] \bigr\rangle.
 \end{array}\right.
\label{eq:StrengF}
\end{equation}
Here (and hereafter if not stated explicitly)
the expectation values are taken with respect to
the cranked mean-field state (the generalized product state)
$|\Phi(\omega_{\rm rot})\rangle$, which is
a vacuum state of the quasiparticles in the uniformly rotating frame
about the $x$-axis,
\begin{equation}
 h-\omega_{\rm rot}J_x = \sum_{\alpha}E_\alpha a^\dagger_\alpha a_\alpha,
\label{eq:crankedQP}
\end{equation}
where $\omega_{\rm rot}$ is the rotational (cranking) frequency,
$E_\alpha$ is the energy of the quasiparticle eigenstate $|\alpha \rangle$
in the rotating frame (quasiparticle routhian),
and $a^\dagger_\alpha$ is its quasiparticle creation operator.
It should be stressed that the vacuum mean-field state,
$|\Phi(\omega_{\rm rot})\rangle$,
determines the strengths in contrast to the conventional case,
where the two-body interaction, such as the $QQ$ force, is given
and it determines the vacuum mean-field state.
This is very important to guarantee the NG mode decoupling:
It is usually a numerically demanding task to satisfy the selfconsistency
between the interaction and the vacuum mean-field state
in such a level as to exactly satisfy the NG mode decoupling.

The creation operator of the RPA eigenmode excited on
the cranked vacuum state $|\Phi(\omega_{\rm rot})\rangle$,
i.e., the one in the uniformly rotating (UR) frame,
is given as a linear combination of the two-quasiparticle excited states,
\begin{equation}
 X^\dagger_n = \sum_{\alpha<\beta}\left[
  \psi_n(\alpha\beta) a^\dagger_\alpha a^\dagger_\beta+
  \phi_n(\alpha\beta) a_\beta a_\alpha \right],
\label{eq:RPAX}
\end{equation}
and is obtained by solving the RPA equation in the UR frame,
\begin{equation}
 [H_{\rm UR}, X^\dagger_n]_{\rm RPA}=\omega_n X^\dagger_n, \quad
 H_{\rm UR}\equiv H-\omega_{\rm rot}J_x,
\label{eq:RPAeqUR}
\end{equation}
where $\omega_n$ is the corresponding RPA eigenfrequency.
Here and hereafter we use $\hbar=1$ unit for simplicity.

The operators which play a key role for defining the principal axis (PA)
frame discussed in the next subsection, are the non-diagonal part
of the (mass) quadrupole tensor operators.
What is relevant to the RPA wobbling theory is their signature $(-)$ part,
\begin{equation}
 \left\{ \begin{array}{l}
{\displaystyle
 Q_y \equiv Q_{21}^{(-)} = -\sqrt{\frac{15}{4\pi}}\sum_{a=1}^A (xz)_a,
}\vspace*{2mm}\\
{\displaystyle
 Q_z \equiv Q_{22}^{(-)} = i\sqrt{\frac{15}{4\pi}}\sum_{a=1}^A (xy)_a,
} \end{array}\right.
\label{eq:defQ}
\end{equation}
where the $z$-axis is chosen as a quantization axis of
the signature classified quadrupole tensor,
\begin{equation}
 Q^{(\pm)}_{2K}=\frac{1}{\sqrt{2(1+\delta_{K0})}}
 \bigl( Q_{2K} \pm Q_{2-K} \bigr),\quad (K=0,1,2).
\label{eq:sigQ}
\end{equation}
They are important also because the out-of-band $E2$ transition
probabilities are calculated by their RPA matrix elements.
The part composed of the two quasiparticle excitations
only contributes within the RPA order, and it
is denoted with the superscript ${(A)}$ in the followings;
\begin{equation}
 \left\{ \begin{array}{l}
{\displaystyle
 (Q_y)^{(A)}= \sum_{\alpha<\beta} q_y(\alpha\beta)
 \bigl( a^\dagger_\alpha a^\dagger_\beta + a_\beta a_\alpha \bigr),
 }\vspace*{1mm}\\
{\displaystyle
 (Q_z)^{(A)}= \sum_{\alpha<\beta} q_z(\alpha\beta)
 \bigl( a^\dagger_\alpha a^\dagger_\beta - a_\beta a_\alpha \bigr),
 }\end{array}\right.
\label{eq:QmatA}
\end{equation}
where we used the phase convention~\cite{BMtext69}
that the two-quasiparticle matrix elements
in this and the following Eq.~(\ref{eq:JmatA}) are all real,
$q_{y,z}(\alpha\beta)= q^*_{y,z}(\alpha\beta)$,
$j_{y,z}(\alpha\beta)= j^*_{y,z}(\alpha\beta)$.
As it can be easily checked,
the signature $(-)$ part of the angular momentum operators,
\begin{equation}
 \left\{ \begin{array}{l}
{\displaystyle
 (iJ_y)^{(A)}= \sum_{\alpha<\beta} j_y(\alpha\beta)
 \bigl( a^\dagger_\alpha a^\dagger_\beta - a_\beta a_\alpha \bigr),
 }\vspace*{1mm}\\
{\displaystyle
 \,\,(J_z)^{(A)}= \sum_{\alpha<\beta} j_z(\alpha\beta)
 \bigl( a^\dagger_\alpha a^\dagger_\beta + a_\beta a_\alpha \bigr),
 }\end{array}\right.
\label{eq:JmatA}
\end{equation}
combines into the RPA eigenmode corresponding to the NG mode
in Eq.~(\ref{eq:RPAeqUR}),
\begin{equation}
 X_{n={\rm NG}}^\dagger
 = \frac{1}{\sqrt{2\langle J_x \rangle}}\left[ (iJ_y)^{(A)}+(J_z)^{(A)}\right],
\label{eq:XNG}
\end{equation}
with the eigenfrequency $\omega_{n={\rm NG}}=\omega_{\rm rot}$.
Here the expectation value of the angular momentum along
the cranking axis can be written as
\begin{equation}
  \langle J_x \rangle
  = 2\sum_{\alpha<\beta}j_y(\alpha\beta)j_z(\alpha\beta).
\label{eq:exptJx}
\end{equation}
In the same way, for the operators $F_{y,z}$ in Eq.~(\ref{eq:defF}),
\begin{equation}
 \left\{ \begin{array}{l}
{\displaystyle
 (F_y)^{(A)}= \sum_{\alpha<\beta} f_y(\alpha\beta)
 \bigl( a^\dagger_\alpha a^\dagger_\beta + a_\beta a_\alpha \bigr),
 }\vspace*{1mm}\\
{\displaystyle
 (F_z)^{(A)}= \sum_{\alpha<\beta} f_z(\alpha\beta)
 \bigl( a^\dagger_\alpha a^\dagger_\beta - a_\beta a_\alpha \bigr),
 }\end{array}\right.
\label{eq:FmatA}
\end{equation}
and by the definition their matrix elements are expressed as
\begin{equation}
 \left\{ \begin{array}{l}
 f_y(\alpha\beta)=E_{\alpha\beta}j_y(\alpha\beta)
  -\omega_{\rm rot}j_z(\alpha\beta),
 \vspace*{1mm}\\
 f_z(\alpha\beta)=E_{\alpha\beta}j_z(\alpha\beta)
  -\omega_{\rm rot}f_y(\alpha\beta),
 \end{array}\right.
\label{eq:FmatAJ}
\end{equation}
with $E_{\alpha\beta}\equiv E_\alpha+E_\beta$,
and the force strengths as
\begin{equation}
 \left\{ \begin{array}{l}
{\displaystyle
 \chi_y^{-1}=2\sum_{\alpha<\beta}E_{\alpha\beta}j^2_y(\alpha\beta)
  -2\omega_{\rm rot}\sum_{\alpha<\beta}j_y(\alpha\beta)j_z(\alpha\beta),
 }\vspace*{1mm}\\
{\displaystyle
 \chi_z^{-1}=2\sum_{\alpha<\beta}E_{\alpha\beta}j^2_z(\alpha\beta)
  -2\omega_{\rm rot}\sum_{\alpha<\beta}j_y(\alpha\beta)j_z(\alpha\beta).
 }\end{array}\right.
\label{eq:StrengFJ}
\end{equation}

Since the residual interaction is of the multi-component separable type,
it is straightforward to solve the RPA equation by means of
the dispersion matrix technique~\cite{EMR80,SM83}.
The forward and backward amplitudes in Eq.~(\ref{eq:RPAX}) are given by
\begin{equation}
 \left\{ \begin{array}{l}
{\displaystyle
 \psi_n(\alpha\beta)=\frac{
  f_y(\alpha\beta)\chi_y {\cal F}_y(n)+f_z(\alpha\beta)\chi_z {\cal F}_z(n)
 }{E_{\alpha\beta}-\omega_n},
 }\vspace*{1mm}\\
{\displaystyle
 \phi_n(\alpha\beta)=\frac{
 -f_y(\alpha\beta)\chi_y {\cal F}_y(n)+f_z(\alpha\beta)\chi_z {\cal F}_z(n)
 }{E_{\alpha\beta}+\omega_n},
 }\end{array}\right.
\label{eq:FBamp}
\end{equation}
where the RPA transition matrix elements
for the operator $F_{y,z}$ are defined by
\begin{equation}
 \left\{ \begin{array}{l}
{\displaystyle
 {\cal F}_y(n)\equiv \bigl\langle [X_n,F_y]\bigr\rangle
 =\sum_{\alpha<\beta}
  \bigl(\psi_n(\alpha\beta) -\phi_n(\alpha\beta)\bigr)f_y(\alpha\beta),
 }\vspace*{1mm}\\
{\displaystyle
 {\cal F}_z(n)\equiv \bigl\langle [X_n,F_z]\bigr\rangle
 =\sum_{\alpha<\beta}
  \bigl(\psi_n(\alpha\beta)+\phi_n(\alpha\beta)\bigr)f_z(\alpha\beta),
 }\end{array}\right.
\label{eq:AmpF}
\end{equation}
which are obtained by solving the following linear equation
\begin{equation}
 \left[ \begin{array}{cc}
  R_{yy}(\omega_n)-\chi_y^{-1}, & R_{yz}(\omega_n)\\
  R_{zy}(\omega_n), & R_{zz}(\omega_n)-\chi_z^{-1}
  \end{array} \right]
 \left[ \begin{array}{l}
  \chi_y{\cal F}_y(n)\\ \chi_z{\cal F}_z(n)
  \end{array} \right]=0.
\label{eq:EqMatF}
\end{equation}
Here the quantities in the matrix are given by
\begin{equation}
 \begin{array}{l}
 R_{yy}(\omega)=D^{(1)}(f_y,f_y;\omega),\,\,
 R_{zz}(\omega)=D^{(1)}(f_z,f_z;\omega),\\
 R_{yz}(\omega)=R_{zy}(\omega)=D^{(2)}(f_y,f_z;\omega),
 \end{array}
\label{eq:dispMatF}
\end{equation}
where the response functions $D^{(1,2)}$ are defined by
\begin{equation}
 \left\{\begin{array}{l}
{\displaystyle
 D^{(1)}(f,g;\omega)\equiv\sum_{\alpha<\beta}
  \frac{2E_{\alpha\beta}f(\alpha\beta)g(\alpha\beta)}
   {E^2_{\alpha\beta}-\omega^2},
 }\vspace*{1mm}\\
{\displaystyle
 D^{(2)}(f,g;\omega)\equiv\sum_{\alpha<\beta}
  \frac{2\omega f(\alpha\beta)g(\alpha\beta)}
   {E^2_{\alpha\beta}-\omega^2}.
 }\end{array} \right.
 \vspace*{2mm}
\label{eq:defdispMat}
\end{equation}
Thus the RPA eigenfrequency $\omega=\omega_n$ is obtained by solving
the dispersion equation
\begin{equation}
 {\rm det}\left| \begin{array}{cc}
  R_{yy}(\omega)-\chi_y^{-1}, & R_{yz}(\omega)\\
  R_{zy}(\omega), & R_{zz}(\omega)-\chi_z^{-1}
  \end{array} \right|=0,
\label{eq:dispEq}
\end{equation}
and the norm of the corresponding solution $(\chi_i{\cal F}_i(n),\,i=y,z)$
is determined by the normalization condition
$\sum_{\alpha<\beta}
\bigl(\psi^2_n(\alpha\beta)-\phi^2_n(\alpha\beta)\bigr)=1$.

In the case of the residual interaction~(\ref{eq:resInt}) with~(\ref{eq:defF}),
the NG mode~(\ref{eq:XNG}) can be explicitly separated out from the equation.
By inserting Eqs.~(\ref{eq:FmatAJ}) and~(\ref{eq:StrengFJ})
into Eq.~(\ref{eq:EqMatF}), the matrix
in Eq.~(\ref{eq:EqMatF}) can be written as
\begin{equation}
 \left[ \begin{array}{cc}
  \omega_n B_y(\omega_n)-\omega_{\rm rot} A_z(\omega_n),&
  \omega_n A_z(\omega_n)-\omega_{\rm rot} B_y(\omega_n)\\
  \omega_n A_y(\omega_n)-\omega_{\rm rot} B_z(\omega_n),&
  \omega_n B_z(\omega_n)-\omega_{\rm rot} A_y(\omega_n)
  \end{array} \right],
\label{eq:MatJ1}
\end{equation}
with the definitions,
\begin{equation}
 \left\{\begin{array}{l}
  A_y(\omega)\equiv \omega_{\rm rot}\bigl({\cal J}_x - {\cal J}_y(\omega)\bigr)
   + \omega {\cal J}_{yz}(\omega),\\
  A_z(\omega)\equiv \omega_{\rm rot}\bigl({\cal J}_x - {\cal J}_z(\omega)\bigr)
   + \omega {\cal J}_{yz}(\omega),\\
  B_y(\omega)\equiv \omega {\cal J}_{y}(\omega)
   - \omega_{\rm rot}{\cal J}_{yz}(\omega),\\
  B_z(\omega)\equiv \omega {\cal J}_{z}(\omega)
   - \omega_{\rm rot}{\cal J}_{yz}(\omega),\\
 \end{array}\right.
\label{eq:defAB}
\end{equation}
and
\begin{equation}
 \begin{array}{c}
{\displaystyle
  {\cal J}_x \equiv \frac{\langle J_x \rangle}{\omega_{\rm rot}},\quad
  {\cal J}_y(\omega) \equiv D^{(1)}(j_y,j_y;\omega),
 }\vspace*{1mm}\\
  {\cal J}_z(\omega) \equiv D^{(1)}(j_z,j_z;\omega),\,\,\,
  {\cal J}_{yz}(\omega) \equiv D^{(2)}(j_y,j_z;\omega).
 \end{array}
\label{eq:defJJ}
\end{equation}
Then Eq.~(\ref{eq:EqMatF}) can be cast into the form
\begin{equation}
 \begin{array}{l}
 \quad\left[ \begin{array}{cc}
 \omega_{\rm rot}-\omega_n,\,\,0\\
 0,\,\,\omega_{\rm rot}+\omega_n
  \end{array} \right] \times
 \vspace*{1mm}\\
 \left[ \begin{array}{cc}
 A_z(\omega_n)+B_y(\omega_n), B_z(\omega_n)+A_y(\omega_n)\\
 A_z(\omega_n)-B_y(\omega_n), B_z(\omega_n)-A_y(\omega_n)
  \end{array} \right]
 \left[ \begin{array}{l}
  \chi_y{\cal F}_y(n)\\ \chi_z{\cal F}_z(n)
  \end{array} \right]=0,
 \end{array}
\label{eq:EqMatJ}
\end{equation}
from which the $\omega_{n={\rm NG}}=\omega_{\rm rot}$ NG solution
is apparent, because by using Eq.~(\ref{eq:XNG}) the NG amplitudes
$(\chi_i{\cal F}_i,\,i=y,z)$ reduce to
\begin{equation}
 \left[ \begin{array}{l}
  \chi_y{\cal F}_y(n={\rm NG})\\ \chi_z{\cal F}_z(n={\rm NG})
  \end{array} \right]=\frac{1}{\sqrt{2\langle J_x \rangle}}
 \left[ \begin{array}{l} 1 \\ 1 \end{array} \right].
\label{eq:NGF}
\end{equation}
Thus, for the non-NG solutions, $\omega_n \ne \omega_{\rm rot}$,
\begin{equation}
 \left[ \begin{array}{cc}
 B_y(\omega_n), A_y(\omega_n) \\
 A_z(\omega_n), B_z(\omega_n)
  \end{array} \right]
 \left[ \begin{array}{l}
  \chi_y{\cal F}_y(n)\\ \chi_z{\cal F}_z(n)
  \end{array} \right]=0,
\label{eq:EqMatJ1}
\end{equation}
and the dispersion equation for them is
\begin{equation}
 {\rm det}\left| \begin{array}{cc}
  B_y(\omega), A_y(\omega) \\
  A_z(\omega), B_z(\omega)
  \end{array} \right|=0,
\label{eq:dispEqJ1}
\end{equation}
which was first derived in Ref.~\cite{JM79}.
Alternatively, Eq.~(\ref{eq:EqMatF}) can be cast into an another form
\begin{equation}
 \left[ \begin{array}{cc}
 A_y(\omega_n), B_z(\omega_n)\\
 B_y(\omega_n), A_z(\omega_n)
  \end{array} \right]
 \left[ \begin{array}{l}
  \bar{\mit\Omega}_y(n)\\
  \bar{\mit\Omega}_z(n)
  \end{array} \right]=0,
\label{eq:EqMatJ2}
\end{equation}
with the definitions,
\begin{equation}
 \left\{ \begin{array}{l}
  \bar{\mit\Omega}_y(n)\equiv
  \omega_n\chi_y{\cal F}_y(n)-\omega_{\rm rot}\chi_z{\cal F}_z(n), \\
  \bar{\mit\Omega}_z(n)\equiv
  \omega_n\chi_z{\cal F}_z(n)-\omega_{\rm rot}\chi_y{\cal F}_y(n).
  \end{array} \right.
\label{eq:AngfrS}
\end{equation}
Again, this equation contains the trivial NG solution~(\ref{eq:NGF}),
i.e., $\bar{\mit\Omega}_y(n)=\bar{\mit\Omega}_z(n)=0$,
with $\omega_{n={\rm NG}}=\omega_{\rm rot}$.
For the non-NG solutions,
$\bar{\mit\Omega}_y(n)/\bar{\mit\Omega}_z(n)
=-B_z(\omega_n)/A_y(\omega_n)
=-A_z(\omega_n)/B_y(\omega_n)$, therefore, by defining
\begin{equation}
 \left\{ \begin{array}{l}
{\displaystyle
  \bar{\cal J}^{\rm eff}_y(n)\equiv
  {\cal J}_y(\omega_n) 
  - {\cal J}_{yz}(\omega_n)\frac{A_y(\omega_n)}{B_z(\omega_n)},
 }\vspace*{1mm}\\
{\displaystyle
  \bar{\cal J}^{\rm eff}_z(n)\equiv
  {\cal J}_z(\omega_n) 
  - {\cal J}_{yz}(\omega_n)\frac{A_z(\omega_n)}{B_y(\omega_n)},
 } \end{array} \right.
\label{eq:effJS}
\end{equation}
Eq.~(\ref{eq:EqMatJ2}) can be further changed into
\begin{equation}
 \left[ \begin{array}{cc}
 \omega_{\rm rot}\bigl({\cal J}_x -\bar{\cal J}^{\rm eff}_y(n)\bigr),
 \,\,\omega_n \bar{\cal J}^{\rm eff}_y(n) \\
 \omega_n \bar{\cal J}^{\rm eff}_z(n),
 \,\,\omega_{\rm rot}\bigl({\cal J}_x -\bar{\cal J}^{\rm eff}_z(n)\bigr)
  \end{array} \right]
 \left[ \begin{array}{l}
  \bar{\mit\Omega}_y(n)\\
  \bar{\mit\Omega}_z(n)
  \end{array} \right]=0,
\label{eq:EqMatJS}
\end{equation}
from which the wobbling phonon energy is formally solved,
\begin{equation}
 \omega_n=\omega_{\rm rot}\sqrt{
 \frac{\bigl({\cal J}_x -\bar{\cal J}^{\rm eff}_y(n)\bigr)
     \bigl({\cal J}_x -\bar{\cal J}^{\rm eff}_z(n)\bigr)}
 {\bar{\cal J}^{\rm eff}_y(n)\bar{\cal J}^{\rm eff}_z(n)}
 }.
\label{eq:WobbEqS}
\end{equation}
This equation is equivalent to Eq.~(\ref{eq:dispEqJ1})
and was first derived by Marshalek~\cite{Mar79}.

In the original formulation of Ref.~\cite{Mar79},
the quantities $\bar{\mit\Omega}_{y,z}(n)$ and
$\bar{\cal J}^{\rm eff}_{y,z}(n)$ are interpreted
as the amplitudes of $y,z$-components of the angular frequency vector
and of moments of inertia, respectively,
in the principal axis (PA) frame, when the $n$-th RPA mode is excited.
However, this is {\it not} the case
for the general deformed mean-field like in the present work,
which is the subject of the next subsection.

For the sake of completeness,
the $E2$ and $M1$ transition probabilities
from the $n$-th RPA phonon excited rotational band
to the vacuum band are calculated,
within the lowest order in the $1/I$ expansion consistent
to the RPA order~\cite{Mar77}, by
\begin{gather}
 B(E2; In \rightarrow I\pm 1{\rm vac})\approx
  \bigl|\bigl\langle [Q^E_{2\pm 1},X_n^\dagger]\bigr\rangle\bigr|^2,
\label{eq:BE2out} \\
 B(M1; In \rightarrow I\pm 1{\rm vac})\approx
  \bigl|\bigl\langle [\mu_{1\pm 1},X_n^\dagger]\bigr\rangle\bigr|^2.
\label{eq:BM1out}
\end{gather}
Here the electric $E2$ and $M1$ operators defined with respect to the $x$-axis,
$Q^E_{2\nu}$ and $\mu_{1\nu}$ ($\nu=\pm 1$) are given by
\begin{gather}
 \begin{array}{ll}
 Q^E_{2\pm 1}
 &={\displaystyle \frac{i}{\sqrt{2}}\bigl(
  Q^{E(-)}_{21} \pm Q^{E(-)}_{22} \bigr)}\\
 &={\displaystyle \frac{i}{\sqrt{2}}\bigl(
  Q^E_y \pm Q^E_z \bigr),} \end{array}
\label{eq:E2out}\\
 \mu_{1\pm 1}=\pm \frac{i}{\sqrt{2}}\bigl(
 i\mu_y \mp \mu_z \bigr),
\label{eq:M1out}
\end{gather}
where the electric quadrupole operators,
$Q^{E(-)}_{2K}$ ($K=1,2$) and $Q^E_{y,z}$, are defined in the same way
as in Eqs.~(\ref{eq:sigQ}) and~(\ref{eq:defQ}),
but only the proton contributions are
included and the electric charge $e$ is multiplied.
The magnetic moment operator $\mu_k$ ($k=x,y,z$) is given as usual,
\begin{equation}
 \mu_k =\sqrt{\frac{3}{4\pi}}\,\mu_N\sum_{a=1}^A
 (g_l^{(\tau)} l_k+g_s^{(\tau)} s_k)_a\quad(\tau=\pi,\nu).
\label{eq:mu}
\end{equation}
As for the in-band stretched $E2$ transition probabilities,
those in the phonon excited band and the vacuum band are the same
in the RPA order and are calculated by
\begin{equation}
\begin{array}{l}
 B(E2; I \rightarrow I\pm 2)\approx
 \frac{1}{2}
  \bigl|\bigl\langle Q^E_{2\pm 2}\bigr\rangle\bigr|^2
 \vspace*{2mm}\\
 \qquad\qquad=\frac{1}{2} \bigl|\bigl\langle
  \bigl(\frac{\sqrt{3}}{2}Q^{E(+)}_{20}+\frac{1}{2}Q^{E(+)}_{22}
   \bigr) \bigr\rangle\bigr|^2.
\end{array}
\label{eq:BE2in}
\end{equation}

\subsection{Interpretation of the RPA eigenmode in the principal axis frame}
\label{sect:IntpPA}

All the calculations of the observable quantities such as excitation
energies and electromagnetic transition probabilities can be done
within the RPA in the UR frame as has been reviewed in the last subsection.
It is, however, necessary to go into the PA frame in order to
interpret the obtained RPA eigenmode and compare its property
with that of the wobbling motion,
which is introduced in the macroscopic rotor model.
Furthermore, the transformation is inevitable
to investigate how much the angular momentum vector,
or the angular frequency vector, tilts or wobbles around the main rotation axis.
This has been done by Marshalek~\cite{Mar79},
and reviewed from a slightly general view point in Ref.~\cite{SM95}
in the light of the quantum theory of
collective coordinates~\cite{BesKur90,GB94}.
In this subsection,
we recapitulate some results from Ref.~\cite{SM95},
in order to show that a slight modification of the original
formulation in~\cite{Mar79}
is necessary for the general deformed mean-field.

The transformation from the laboratory frame to the UR frame is an
unitary transformation so that the RPA in the UR frame satisfies all
the usual properties~\cite{RingSchuck}, e.g., the completeness
relation.  Therefore, the two-quasiparticle excitation part of
the quadrupole operators in Eq.~(\ref{eq:QmatA}) can be expanded
in terms of the RPA eigenmodes,
\begin{equation}
 \left\{ \begin{array}{l}
{\displaystyle
 (Q_y)^{(A)}= \sum_{n:{\rm all}} {\cal Q}_y(n)
 \bigl( X_n^\dagger + X_n \bigr),
 }\vspace*{1mm}\\
{\displaystyle
 (Q_z)^{(A)}= \sum_{n:{\rm all}} {\cal Q}_z(n)
 \bigl( X_n^\dagger - X_n \bigr),
 }\end{array}\right.
\label{eq:QmatX}
\end{equation}
where the amplitudes ${\cal Q}_{y,z}(n)$ are calculated by
\begin{equation}
 \left\{ \begin{array}{l}
{\displaystyle
 {\cal Q}_y(n)\equiv\bigl\langle [X_n,Q_y] \bigr\rangle
  = \sum_{\alpha<\beta}\bigl(\psi_n(\alpha\beta)-\phi_n(\alpha\beta)
   \bigr)q_y(\alpha\beta),
 }\vspace*{1mm}\\
{\displaystyle
 {\cal Q}_z(n)\equiv\bigl\langle [X_n,Q_z] \bigr\rangle
  = \sum_{\alpha<\beta}\bigl(\psi_n(\alpha\beta)+\phi_n(\alpha\beta)
   \bigr)q_z(\alpha\beta).
 }\end{array}\right.
\label{eq:Qamp}
\end{equation}
In particular, the NG mode contribution is contained in Eq.~(\ref{eq:QmatX}),
\begin{equation}
 \left\{ \begin{array}{l}
{\displaystyle
 {\cal Q}_y(n={\rm NG})=-\frac{2\alpha_y R^2}{\sqrt{2\langle J_x \rangle}},
 }\vspace*{1mm}\\
{\displaystyle
 {\cal Q}_z(n={\rm NG})= \frac{2\alpha_z R^2}{\sqrt{2\langle J_x \rangle}},
 }\end{array}\right.
\label{eq:QampNG}
\end{equation}
where $R^2\equiv \langle\, \sum_{a=1}^A (\mbox{\boldmath$r$}^2)_a \rangle$
is the mean square radius and $\alpha_y$($\alpha_z$) describes the
deformation around the $y$\,($z$)-axis,
\begin{equation}
 \left\{ \begin{array}{l}
 \begin{array}{l}
{\displaystyle
 2\alpha_y R^2 \equiv \sqrt{\frac{15}{4\pi}} \bigl\langle\,
  \sum_{a=1}^A(x^2-z^2)_a \bigr\rangle
  }\\
{\displaystyle
  \qquad= \bigl\langle [iJ_y,Q_y]  \bigr\rangle
  =2\sum_{\alpha<\beta}j_y(\alpha\beta)q_y(\alpha\beta),
 }\end{array} \vspace*{1mm}\\
 \begin{array}{l}
{\displaystyle
 2\alpha_z R^2 \equiv \sqrt{\frac{15}{4\pi}} \bigl\langle\,
  \sum_{a=1}^A(x^2-y^2)_a \bigr\rangle
  }\\
{\displaystyle
  \qquad= \bigl\langle [J_z,Q_z]  \bigr\rangle
  =2\sum_{\alpha<\beta}j_z(\alpha\beta)q_z(\alpha\beta).
 }\end{array} \end{array}\right.
\label{eq:defP}
\end{equation}
In contrast, the two-quasiparticle excitation part
of the angular momentum operators are composed purely of the NG mode,
\begin{equation}
 \left\{ \begin{array}{l}
{\displaystyle
 (iJ_y)^{(A)}= \sqrt{\frac{\langle J_x \rangle}{2}}
 \bigl(X_{\rm NG}^\dagger - X_{\rm NG}\bigr),
 }\vspace*{1mm}\\
{\displaystyle
 \,\,(J_z)^{(A)}= \sqrt{\frac{\langle J_x \rangle}{2}}
 \bigl(X_{\rm NG}^\dagger + X_{\rm NG}\bigr).
 }\end{array}\right.
\label{eq:JmatX}
\end{equation}

Now let us consider the behavior of
the angular momentum vector and the angular frequency vector
when the RPA wobbling mode is excited,
from which the moments of inertia are naturally introduced.
We do not go into the full details of the theory,
but briefly mention the procedure.
As it is well-known, the (three dimensional) cranking prescription
can be viewed as a result of the constraints, $J_i=I_i$ ($i=x,y,z$),
where $I_i$ are time-dependent constants in the semiclassical
theory~\cite{KO81,Kaneko94}, or are the total angular momentum operators
in the rotor model (i.e., the differential operators with respect
to the Euler angles) in the quantum theory of collective
coordinates~\cite{KBC90}.  This kind of constraints are called
the first class constraints in the general theory~\cite{Dirac64},
and one has to impose the second class constraints,
i.e., the PA frame condition in this case,
in order to develop the consistent framework of the quantum~\cite{BesKur90}
or the classical canonical~\cite{SCC-YK} theory.
The second class constraints can be, in principle, arbitrarily chosen
(the gauge condition), and the physical observables should not depend
on the choice.  However, the actual choice is important for the
physical picture; recall, e.g., the choice of the center of mass coordinate
in the case of the translational motion.
Here we take the following PA frame condition~\cite{VC70,Mar79,KO81};
the non-diagonal part of the quadrupole tensor should vanish identically:
\begin{equation}
 (Q_y)_{\rm PA}\equiv 0,\quad
 (Q_z)_{\rm PA}\equiv 0,
\label{eq:condPA}
\end{equation}
(the constraint for the $x$-component, i.e.,
$(Q_x)_{\rm PA}\equiv (Q_{21}^{(+)})_{\rm PA}\equiv 0$,
is automatically satisfied
in the RPA order),
where the subscript PA denotes that the quantity is expressed
in the PA frame.

This condition relates the collective coordinates, i.e., the Euler angles
in the present case, to the nucleon degrees of freedom.
Although it is a difficult task in general,
it has been shown in Ref.~\cite{Mar79} that this relation can be
solved within the small amplitude limit (RPA).
The UR frame is transformed to the PA frame by
the rotation in terms of the dynamical Euler angle variables $\Theta$,
\begin{equation}
 ({\cal O})_{\rm PA} = U(\Theta){\cal O}U^{-1}(\Theta),
\label{eq:TrPA}
\end{equation}
for an arbitrary operator ${\cal O}$. Here we omit the subscript ${\rm UR}$
for the quantity in the UR frame.  Within the RPA order~\cite{Mar79},
\begin{equation}
 U(\Theta) \approx 1 - i\theta\bigl(\sin\psi J_y + \cos\psi J_z\bigr),
\label{eq:UTrPA}
\end{equation}
where $\Theta=(\psi,\theta,\phi)$ are Euler angles with respect to
the $x$-axis (see Ref.~\cite{Mar79}),
and the small amplitude limit, $\theta \ll 1$, is used.
From Eqs.~(\ref{eq:UTrPA}) and~(\ref{eq:TrPA}),
\begin{equation}
 ({\cal O})_{\rm PA} \approx {\cal O}
   + \theta\sin\psi \bigl\langle [{\cal O},iJ_y] \bigr\rangle
   + i\theta\cos\psi \bigl\langle [{\cal O},J_z] \bigr\rangle,
\label{eq:TrRPA}
\end{equation}
and applying this to Eq.~(\ref{eq:condPA}),
one obtains
\begin{equation}
 \theta\sin\psi = \frac{Q_y}{2\alpha_y R^2},\quad
 i\theta\cos\psi = \frac{Q_z}{2\alpha_z R^2}.
\label{eq:AngQ}
\end{equation}
In this way the collective Euler-angle variables can be expressed directly
by the nucleon degrees of freedom within the RPA order,
i.e., the constraint conditions are approximately solved
(the remaining angle $\phi$ does not play any role within the RPA order).
Applying the transformation~(\ref{eq:TrRPA}) to the angular momentum operators,
\begin{equation}
 \left\{ \begin{array}{l}
 \begin{array}{l}
{\displaystyle
 (iJ_y)^{(A)}_{\rm PA}=\Bigl(iJ_y
 -\frac{\langle J_x \rangle}{2\alpha_zR^2}Q_z\Bigr)^{(A)}
 } \vspace*{1mm}\\
{\displaystyle \qquad\qquad
 =-\langle J_x \rangle
 \sum_{n\ne{\rm NG}} r_z(n) \bigl(X_n^\dagger - X_n \bigr),
 }\end{array} \vspace*{1mm}\\
 \begin{array}{l}
{\displaystyle
 (J_z)^{(A)}_{\rm PA}=\Bigl(J_z
 +\frac{\langle J_x \rangle}{2\alpha_yR^2}Q_y\Bigr)^{(A)}
 } \vspace*{1mm}\\
{\displaystyle \qquad\qquad
 = \langle J_x \rangle
 \sum_{n\ne{\rm NG}} r_y(n) \bigl(X_n^\dagger + X_n \bigr),
 }\end{array} \end{array}\right.
\label{eq:JPA}
\end{equation}
namely, their amplitudes associated with the $n$-th
RPA eigenmode in the PA frame are
\begin{equation}
 \left\{ \begin{array}{l}
 (J_y)_{\rm PA}(n)\equiv
 \bigl\langle [X_n,(iJ_y)_{\rm PA}] \bigr\rangle
 =-\langle J_x \rangle\,r_z(n),
 \vspace*{1mm}\\
  (J_z)_{\rm PA}(n)\equiv
 \bigl\langle [X_n,(J_z)_{\rm PA}] \bigr\rangle
 = \langle J_x \rangle\,r_y(n),
 \end{array} \right.
\label{eq:ampJPA}
\end{equation}
with the definitions,
\begin{equation}
{\displaystyle
 r_y(n)\equiv \frac{{\cal Q}_y(n)}{2\alpha_y R^2},\quad
 r_z(n)\equiv \frac{{\cal Q}_z(n)}{2\alpha_z R^2},
 }
\label{eq:rQPA}
\end{equation}
which describe the shape fluctuations (ratios of
the dynamic fluctuations amplitude to the static deformations).
Thus, the contribution of the NG mode, which will be identified
as the collective coordinate, is driven out and only the intrinsic
degrees of freedom remain in the PA frame.
From Eq.~(\ref{eq:JPA}) it is clear that
the angular momentum vector fluctuates or wobbles about
the main rotation axis ($x$-axis),
when the $n$-th RPA phonon is excited.
The quantity $r_{y,z}(n)$ represent their fluctuations;
\begin{equation}
{\displaystyle
 r_y(n)=  \frac{(J_z)_{\rm PA}(n)}{\langle J_x \rangle},\quad
 r_z(n)= -\frac{(J_y)_{\rm PA}(n)}{\langle J_x \rangle}.
 }
\label{eq:rJPA}
\end{equation}
Therefore these quantities should be small $r_{y,z}(n) \ll 1$
in order for the RPA to be valid.
Note that all the amplitudes ${\cal Q}_{y,z}(n)$ and $r_{y,z}(n)$
are real because of the present phase convention.

The angular frequency vector operators $({\mit\Omega}_i,i=x,y,z)$
in the PA frame are introduced
as Lagrange multipliers of the first class constraints,
$J_i=I_i$, and
the dynamical time dependence of nucleon operators in the PA frame
is generated by the hamiltonian in the PA frame,
\begin{equation}
 \begin{array}{l}
 H_{\rm PA} = H
 - {\mit\Omega}_x J_x - {\mit\Omega}_y J_y - {\mit\Omega}_z J_z \\
 \qquad \approx H_{\rm UR}  - {\mit\Omega}_y J_y - {\mit\Omega}_z J_z,
 \end{array}
\label{eq:HPA}
\end{equation}
within the RPA order, because ${\mit\Omega}_x \approx \omega_{\rm rot}$.
The angular frequency variables $({\mit\Omega}_{y,z})$ are determined by
the consistency condition of the PA frame (gauge) condition,
i.e., Eq.~(\ref{eq:condPA}) should hold in arbitrary time,
\begin{equation}
 i\frac{d}{dt}(Q_{y,z})_{\rm PA}=\bigl([Q_{y,z},H_{\rm PA}]\bigr)_{\rm PA}=0.
\label{eq:condFreqPA}
\end{equation}
Here
$\bigl([Q_y,H_{\rm PA}]\bigr)_{\rm PA}
\approx \bigl([Q_y,H_{\rm UR}]\bigr)_{\rm PA}
+i{\mit\Omega}_y \bigl\langle [Q_y,iJ_y] \bigr\rangle$,
$\bigl([Q_z,H_{\rm PA}]\bigr)_{\rm PA}
\approx \bigl([Q_z,H_{\rm UR}]\bigr)_{\rm PA}
-{\mit\Omega}_z \bigl\langle [Q_z,J_z] \bigr\rangle$,
and
$H_{\rm UR} \approx
\sum_{n:{\rm all}} \omega_n (X_n^\dagger X_n+\frac{1}{2})$
within the RPA order, and using the transformation~(\ref{eq:TrRPA})
for $\bigl([Q_{y,z},H_{\rm UR}]\bigr)_{\rm PA}$, we obtain
\begin{equation}
 \left\{ \begin{array}{l}
{\displaystyle
 (i{\mit\Omega}_y)^{(A)}
 =-\sum_{n\ne{\rm NG}}
 \bigl(\omega_n r_y(n)+\omega_{\rm rot}r_z(n)\bigr)
  \bigl( X_n^\dagger - X_n \bigr),
 } \vspace*{1mm}\\
{\displaystyle
 ({\mit\Omega}_z)^{(A)}
 = \sum_{n\ne{\rm NG}}
 \bigl(\omega_n r_z(n)+\omega_{\rm rot}r_y(n)\bigr)
  \bigl( X_n^\dagger + X_n \bigr),
 }\end{array} \right.
\label{eq:OmPA}
\end{equation}
i.e., their amplitudes associated with the $n$-th RPA eigenmode are
\begin{equation}
 \left\{ \begin{array}{l}
 {\mit\Omega}_y(n)\equiv
 \bigl\langle [X_n,i{\mit\Omega}_y] \bigr\rangle
 =-\bigl(\omega_n r_y(n)+\omega_{\rm rot}r_z(n)\bigr),
 \vspace*{1mm}\\
 {\mit\Omega}_z(n)\equiv
 \bigl\langle [X_n,{\mit\Omega}_z] \bigr\rangle
 = \omega_n r_z(n)+\omega_{\rm rot}r_y(n).
 \end{array} \right.
\label{eq:Angfr}
\end{equation}
Again, the NG mode contribution cancels out completely.
From Eq.~(\ref{eq:OmPA}) the angular frequency vector
also fluctuates about the main rotation axis,
and the moments of inertia in the PA frame are naturally
introduced;
\begin{equation}
 \left\{ \begin{array}{l}
{\displaystyle
 {\cal J}^{\rm eff}_y(n)\equiv\frac{(J_y)_{\rm PA}(n)}{{\mit\Omega}_y(n)}
 =\frac{{\cal J}_x\omega_{\rm rot}r_z(n)}
  {\omega_n r_y(n)+\omega_{\rm rot} r_z(n)},
 } \vspace*{1mm}\\
{\displaystyle
 {\cal J}^{\rm eff}_z(n)\equiv\frac{(J_z)_{\rm PA}(n)}{{\mit\Omega}_z(n)}
 =\frac{{\cal J}_x\omega_{\rm rot}r_y(n)}
  {\omega_n r_z(n)+\omega_{\rm rot} r_y(n)}.
 }\end{array} \right.
\label{eq:effJ}
\end{equation}
It is clear that the moments of inertia thus
defined are intimately connected to the PA frame condition~(\ref{eq:condPA})
and determined by the RPA transition amplitudes
of the $Q_{y,z}$ operators in Eq.~(\ref{eq:Qamp}).

From Eq.~(\ref{eq:effJ}) the RPA eigenenergy and
the $r_{y,z}(n)$ amplitudes can be represented in terms of
these moments of inertia~\cite{Mar79,SM95};
\begin{equation}
 \omega_n=\omega_{\rm rot}\sqrt{
 \frac{\bigl({\cal J}_x -{\cal J}^{\rm eff}_y(n)\bigr)
     \bigl({\cal J}_x -{\cal J}^{\rm eff}_z(n)\bigr)}
  {{\cal J}^{\rm eff}_y(n){\cal J}^{\rm eff}_z(n)}
 },
\label{eq:WobbEq}
\end{equation}
i.e., the well-known formula in the rotor model~\cite{BMtext75} is
recovered, and
\begin{equation}
 \left\{ \begin{array}{l}
{\displaystyle
 r_y(n)=\,\,c_n\frac{1}{\sqrt{2\langle J_x \rangle}}
  \left[ \frac{1/{\cal J}^{\rm eff}_y(n)-1/{\cal J}_x}
              {1/{\cal J}^{\rm eff}_z(n)-1/{\cal J}_x} \right]^{1/4},
 } \vspace*{1mm}\\
{\displaystyle
 r_z(n)=\sigma_n c_n\frac{1}{\sqrt{2\langle J_x \rangle}}
  \left[ \frac{1/{\cal J}^{\rm eff}_z(n)-1/{\cal J}_x}
              {1/{\cal J}^{\rm eff}_y(n)-1/{\cal J}_x} \right]^{1/4},
 }\end{array} \right.
\label{eq:ampr3J}
\end{equation}
where the quantity $c_n$ and the sign $\sigma_n$ are defined by
\begin{equation}
 \begin{array}{l}
 c_n =  [\mbox{sign of }r_y(n)]\times
 \sqrt{2\langle J_x \rangle |r_y(n) r_z(n)|},\\
 \sigma_n = \mbox{sign of }\bigl(r_y(n) r_z(n)\bigr).
 \end{array}
 \vspace*{1mm}
\label{eq:sigmac}
\end{equation}
Combining Eqs.~(\ref{eq:ampr3J}),~(\ref{eq:rQPA}) and~(\ref{eq:BE2out})
with $Q^E_{2\nu}$ being replaced to $(eZ/A)Q_{2\nu}$, the rotor model
expression for the out-of-band $B(E2)$ is also recovered,
if $\sigma_n=+$ and $c_n^2=1$
are satisfied for the excitation of the $n$-th RPA eigenmode.
This was first shown in Ref.~\cite{SM95}, and the condition,
$\sigma_n c_n^2=2\langle J_x \rangle\,r_y(n) r_z(n)=+1$, is used to identify
the wobbling mode in a recent publication~\cite{KN07}:
Note that $\sigma_n=-$ and $c_n=-1$ for the NG mode.

It must be emphasized that the discussion above is general
and does not depend on specific choices either of residual interactions
or of nuclear mean-fields.
In the original formulation of Marshalek~\cite{Mar79},
the arbitrary spherical mean-field and
the $QQ$ (plus the monopole pairing) force are assumed.
It can be easily checked that the formulation is still valid
for the selfconsistent anisotropic harmonic oscillator potential
model~\cite{BMtext75}, for which the induced residual interaction
is the doubly-stretched $Q''Q''$ force~\cite{Kis75,Sak89,Suz77,Mar84}.
In such cases, the deformation is purely of quadrupole type,
and the selfconsistent mean-field $h$ leads for $F_{y,z}$
in Eq.~(\ref{eq:defF}) to
\begin{equation}
{\displaystyle
 \chi_y F_y= -\frac{1}{2\alpha_y R^2}\,Q_y,\quad
 \chi_z F_z=  \frac{1}{2\alpha_z R^2}\,Q_z,
 }
\label{eq:specF}
\end{equation}
and the quantities $\bar{\mit\Omega}_{y,z}(n)$
and $\bar{\cal J}^{\rm eff}_{y,z}(n)$ introduced
in Eqs.~(\ref{eq:AngfrS}) and~(\ref{eq:effJS}) coincide
with ${\mit\Omega}_{y,z}(n)$ and ${\cal J}^{\rm eff}_{y,z}(n)$
in Eqs.~(\ref{eq:Angfr}) and~(\ref{eq:effJ}), respectively.
However, this is not the case generally,
and the moments of inertia should not be
calculated by Eq.~(\ref{eq:effJS}) but by Eq.~(\ref{eq:effJ})
for the general mean-field: This is the point,
in which the original formulation of Marshalek should be modified.
As is discussed above, the moments of inertia
are determined by the PA frame condition.
Therefore, a possible alternative choice of the PA frame condition is
$(F_{y,z})_{\rm PA}\equiv 0$ in place of $(Q_{y,z})_{\rm PA}\equiv 0$.
Then the original theory is recovered, but now the rotor model
expression for the out-of-band $B(E2)$ cannot be derived,
because the operators $F_{y,z}$ are not related strictly
to the quadrupole operators $Q_{y,z}$ for the general mean-field.
We use Eq.~(\ref{eq:condPA}) as the PA frame condition
for the realistic calculations presented in Sec.~\ref{sect:result}.

\subsection{Remarks on the NG mode decoupling}
\label{sect:NGdecoupl}

In the realistic calculations, it is not an easy task to exactly
realize the NG mode decoupling, even though the microscopic hamiltonian
satisfies the decoupling condition~(\ref{eq:NGdecouple}).
For example, one has to use a truncation of the model space
for diagonalization or a finite mesh size to solve differential
equations of the single-particle states,
which inevitably breaks the condition.
When starting from some effective two-body force and the mean-field
is determined by the HF (or HFB) procedure,
a very accurate achievement of the selfconsistency between
the interaction and the resultant mean-field is required.
Practically, it often happens that even
the decoupling condition~(\ref{eq:NGdecouple}) does not rigorously meet.
For example, if one takes the Nilsson potential and the $QQ$ force
(or the doubly-stretched $Q''Q''$ force) as the total hamiltonian,
then the rotational NG modes do not decouple exactly, because
the general Nilsson potential~\cite{NR95} contains
the ``hexadecapole'' deformed potential (the $\epsilon_4$ deformation),
whose effects cannot be dealt with the simple $QQ$ (or $Q''Q''$) force.
Even if the $\epsilon_4$ deformation is neglected,
the $\mbox{\boldmath$l$}\cdot\mbox{\boldmath$s$}$ and
$\mbox{\boldmath$l$}^2$ terms are those defined in the stretched coordinate
in the standard Nilsson potential,
so that they are not strictly spherically invariant.

The previous calculations of the TSD wobbling excitations
in Refs.~\cite{MSMwob1,MSMwob2,MSMwob3} suffer from this problem;
even though the $\epsilon_4$ deformation was neglected in them,
the stretched $\mbox{\boldmath$l$}\cdot\mbox{\boldmath$s$}$ and
$\mbox{\boldmath$l$}^2$ terms are used.
Then $\bar{\cal J}^{\rm eff}_{y,z}(n)$ in Eq.~(\ref{eq:effJS})
does not coincide with ${\cal J}^{\rm eff}_{y,z}(n)$
in Eq.~(\ref{eq:effJ}) rigorously.
However, we have checked that the effect of the model space truncation
is much severe, and the effect of
the stretched $\mbox{\boldmath$l$}\cdot\mbox{\boldmath$s$}$ and
$\mbox{\boldmath$l$}^2$ terms is very small;
$\bar{\cal J}^{\rm eff}_{y,z}(n)$ and ${\cal J}^{\rm eff}_{y,z}(n)$
agree within 1\% if enough numbers of the oscillator shells are included.
In the recent calculations in Refs.~\cite{KN06,KN07},
the non-stretched $\mbox{\boldmath$l$}\cdot\mbox{\boldmath$s$}$ and
$\mbox{\boldmath$l$}^2$ terms are used in the Nilsson potential
without the $\epsilon_4$ deformation consistently,
but the additional correction term, which recovers the Galilean invariance
broken by the velocity dependent part, breaks the spherical symmetry.
Therefore, in all these calculations, even if $\epsilon_4=0$,
the NG mode decoupling was not strictly realized.
In Refs.~\cite{KN06,KN07}, one of the strengths of the $Q''Q''$ force
is adjusted so that the RPA spectra contain the eigenenergy
$\omega_n=\omega_{\rm rot}$,
but then the corresponding eigenmode solution
might not generally be the correct NG mode, Eq.~(\ref{eq:XNG}),
although it is claimed that such effect is small~\cite{KN06,KN07}.

If the monopole (or seniority) pairing correlation is included
in the mean-field, it can be an another source that disturbs
the actual NG mode decoupling.
Since the monopole (or seniority) pairing interaction gives
divergent results in the full model space,
some truncation scheme is necessary.
Usually the energy truncation of the single-particle states
is used, in which not all of the magnetic substates ($m$)
for the orbital with a given angular momentum ($j$)
are included in the pairing model space,
and then the rotational invariance is broken.
We have checked that this effect is very small
if the pairing model space is not too small.

The method to construct the residual interaction that
restores the rotational symmetry in Sec.~\ref{sect:ResIntUR} can be used
to remedy (or avoid) this problem generally.
Since the wobbling motion is intimately
related to the rotational NG mode, we believe that it is important
to respect its decoupling, and the method in Sec.~\ref{sect:ResIntUR}
is a simple and flexible way to realize it.

\section{Results of numerical calculations}
\label{sect:result}

The purpose of the present work is to study the wobbling excitations
of the triaxial superdeformed (TSD) bands in the Lu and Hf region.
We first discuss the result of the mean-field calculations
based on the new parameterization of the Woods-Saxon potential
explained in Sec.~\ref{sect:WSpot},
and then the result of the microscopic RPA calculations.

\subsection{Mean-field calculations}
\label{sect:calMF}

Our mean-field hamiltonian is
\begin{equation}
 h_\tau=t_\tau+V_{{\rm WS},\tau}(\mbox{\boldmath$r$})
  +\delta_{\tau,\pi}V_{\rm Coul}(\mbox{\boldmath$r$})
  -{\mit\Delta}_\tau(P_\tau^\dagger+P_\tau) -\lambda_\tau N_\tau,
\label{eq:mfh}
\end{equation}
where $\tau=\pi,\nu$ distinguish the proton or neutron part,
$t$ is the kinetic energy term, $V_{{\rm WS},\tau}(\mbox{\boldmath$r$})$
is the Woods-Saxon potential in Eq.~(\ref{eq:WSpot}),
$V_{\rm Coul}(\mbox{\boldmath$r$})$
is the Coulomb potential in Eq.~(\ref{eq:Coulpot}),
$P^\dagger$ and ${\mit\Delta}$
are the monopole pair transfer operator and pairing gap,
and $N$ and $\lambda$ are the number operator and the chemical potential,
respectively.
We have developed our own code, based on the program of Ref.~\cite{CDN87},
for the cranked Woods-Saxon calculation with pairing correlations included.
The diagonalization is performed in two steps; first the Woods-Saxon
(and Coulomb) potential (${\mit\Delta}=0$ and $\omega_{\rm rot}=0$)
is diagonalized in the anisotropic harmonic oscillator basis, and then
the cranked pairing problem (the HFB equation) is solved
in the Woods-Saxon basis obtained in the first step.

The matrix elements of the Woods-Saxon (and Coulomb) potential
with respect to the anisotropic oscillator basis are specified
by the oscillator quantum numbers, $(n_x,n_y,n_z)$.  They are evaluated
by using the recurrence relations of the Hermite polynomials~\cite{CDN87}
starting from the diagonal and subdiagonal matrix elements,
which are calculated by the Gauss-Hermite quadrature formula.
The three oscillator frequencies, $(\omega_x,\omega_y,\omega_z)$,
are chosen to be
inversely proportional to the mean radii of the $x,y,z$-directions,
which are calculated geometrically by using the uniform density distribution
based on the shape specified in Eq.~(\ref{eq:defRS}) for given deformation
parameters ($\beta_2,\gamma,\beta_4$).  The magnitude of the frequencies,
$(\omega_x\omega_y\omega_z)^{1/3}$, is chosen to be $1.25\times \omega_0$,
with $\hbar \omega_0=41/A^{1/3}$ MeV.
We use the truncation of the oscillator shells,
$N_{\rm max}\equiv (n_x+n_y+n_z)_{\rm max}=12$,
in most of the calculations presented in this work.
As for the Coulomb potential~(\ref{eq:Coulpot}),
it is evaluated by the Gauss-Legendre numerical integration if
${\rm min}_{\mit\Omega}R({\mit\Omega}) < r
< {\rm max}_{\mit\Omega}R({\mit\Omega})$, and by using the
multipole expansion of $|\mbox{\boldmath$r$}-\mbox{\boldmath$r$}'|^{-1}$
up to $l_{\rm max}=16$ if otherwise.
Since we use the monopole pairing interaction, the truncation
of the pairing model space is necessary.
Woods-Saxon single-particle orbitals whose energy $\epsilon_i$
satisfies $|\epsilon_i -\lambda| \le \hbar\omega_0$
are included in the pairing model space.

The main reason why we use the Woods-Saxon potential rather than
the Nilsson potential is that the cranking moment of inertia ${\cal J}_x$
in Eq.~(\ref{eq:defJJ}) is overestimated unless
the Strutinsky renormalization~\cite{NR95}
is not performed for the expectation value $\langle J_x \rangle$.
In order to show how this overestimation is greatly improved,
we compare the cranking moment of inertia calculated by using
the Woods-Saxon and the Nilsson potentials
in Table~\ref{tab:crmom}.
The cranking moment of inertia is sensitive to the shell structure
and the pairing correlation, so that we tabulated the Strutinsky
smoothed moment of inertia,
i.e., $\widetilde{\cal J}_x\equiv
\widetilde{\langle J_x \rangle}/\omega_{\rm rot}$, without pairing
for selected nuclei and deformations in the rare earth region.
This inertia $\widetilde{\cal J}_x$ is $\omega_{\rm rot}$-dependent
but its dependence is known to be weak~\cite{ALL76}
(we have checked that the variations are less than 2\%
in the present examples),
so that those at $\omega_{\rm rot}=0.6$ MeV/$\hbar$ are tabulated.
We also include the corresponding rigid-body values calculated
by assuming the uniform density distribution with
the radius $1.2 \times A^{1/3}$ fm.
For the calculations of this $\widetilde{\cal J}_x$
and of the selfconsistent deformation of the TSD minima discussed
in the followings, we employ the cranked Nilsson Strutinsky calculation
of Ref.~\cite{SVB90}, where the ${\mit\Delta}N_{\rm osc}=\pm 2$ coupling
arising from the cranking term is included in an approximate way.
The parameters of the present Woods-Saxon potential are given
in Table~\ref{tab:WSparam}, and those of the Nilsson potential
are taken from Ref.~\cite{BR85}.
The parametrizations of shape in the two potentials are different.
We convert those of the Nilsson potential
($\epsilon_2,\gamma,\epsilon_4$)~\cite{BR85} to
those of the Woods-Saxon potential
($\beta_2,\gamma,\beta_4$) by requiring
the same mean square radii of the $x,y,z$-directions,
which are calculated geometrically by the uniform density distribution.
We use this conversion between 
($\beta_2,\gamma,\beta_4$) and ($\epsilon_2,\gamma,\epsilon_4$)
throughout this work.
As it is clear in Table~\ref{tab:crmom},
the overestimation with the Nilsson potential by 25$-$27\%
is much more reduced to 6$-$7\%; the remaining deviations
are mainly due to the spin-orbit potential.

\begin{table}[hbtp]
\caption{ The Strutinsky smoothed cranking moment of inertia
in unit of $\hbar^2$/MeV for various nuclei and deformations.
The corresponding rigid-body values are also included.
The deformation parameters $(\beta_2,\gamma,\beta_4)$ are
those used in Eq.~(\ref{eq:WSbeta}).
}
\label{tab:crmom}
\begin{ruledtabular}
\begin{tabular}{ccccccc}
nucleus & $\beta_2$ & $\gamma$ & $\beta_4$ &
 $\widetilde{\cal J}_x({\rm WS})$ &
 $\widetilde{\cal J}_x({\rm Nils})$ &
 ${\cal J}_x({\rm rigid})$
 \\
 \hline
$^{164}$Er & 0.258 & $ 0^\circ$ & 0.004 & 82.2 &  96.6 & 77.4 \\
$^{174}$Hf & 0.288 & $ 0^\circ$ &\hspace*{-2.7mm}$-$0.027 & 92.0 & 108.1 & 86.9 \\
$^{152}$Dy & 0.665 & $ 0^\circ$ & 0.134 & 97.1 & 112.6 & 90.6 \\
$^{163}$Lu & 0.420 & $18^\circ$ & 0.020 & 90.9 & 109.0 & 85.7 \\
\end{tabular}
\end{ruledtabular}
\end{table}

The result of Table~\ref{tab:crmom} indicates that
the single-particle orbitals of
the present Woods-Saxon potential have correct radial size,
which is also apparent from the values of the radius parameters
in Table~\ref{tab:WSparam}.
In fact we have checked the proton and neutron density distributions 
calculated by using the present Woods-Saxon potential
are quite similar to those obtained by the Hartree-Fock
calculations using the standard Skyrme and Gogny forces.

\begin{figure*}[htbp]
\includegraphics[angle=0,width=130mm,keepaspectratio]{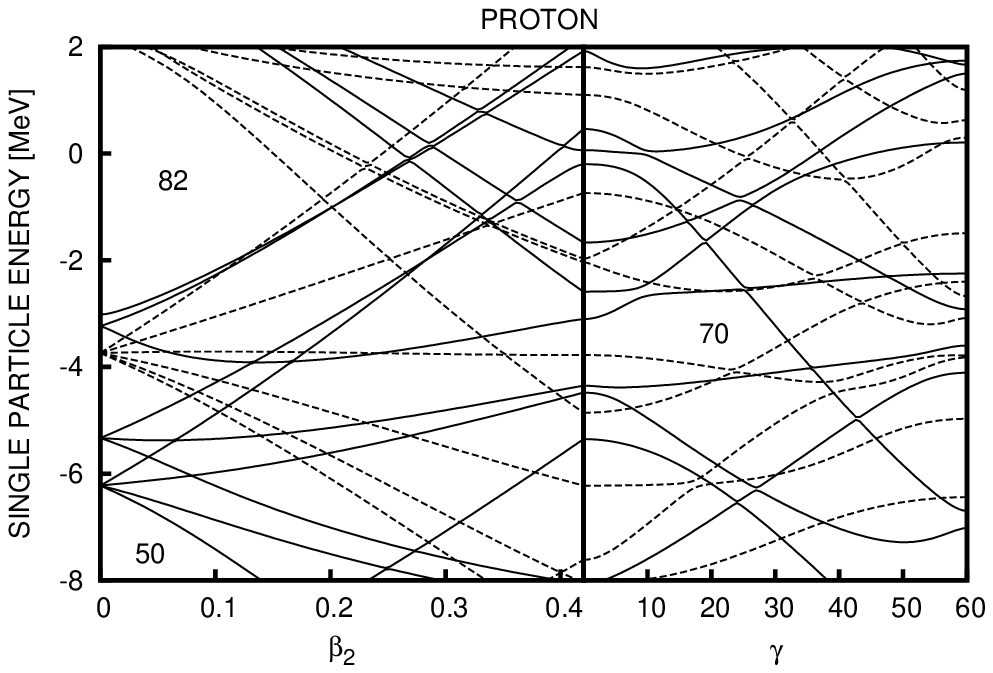}
\includegraphics[angle=0,width=130mm,keepaspectratio]{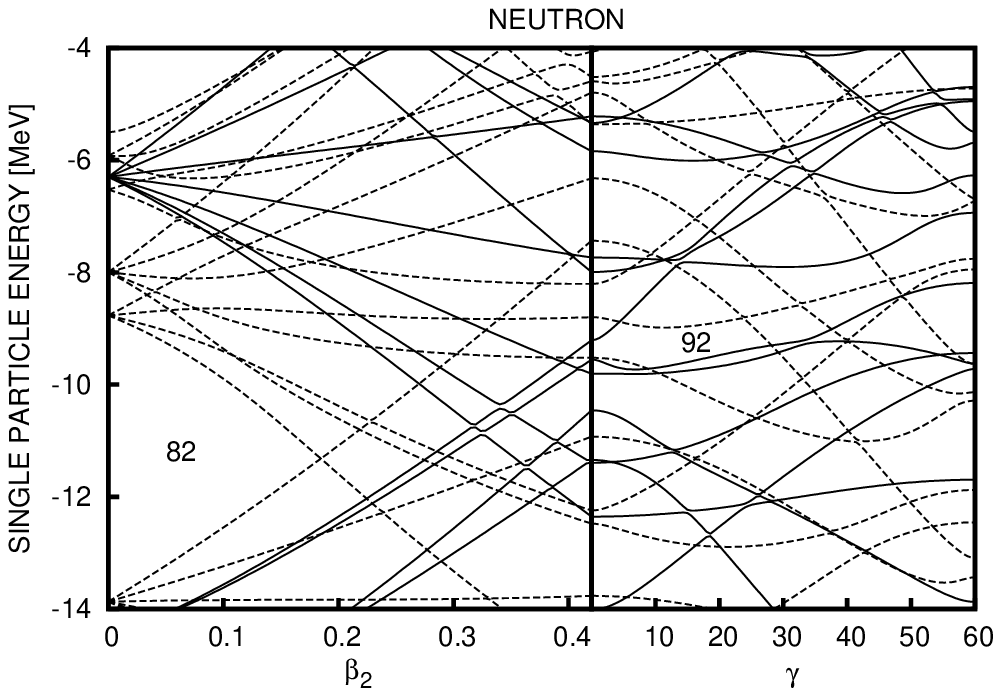}
\caption{
 The single-particle energies as functions of
 $\beta_2(\beta_4)$ and $\gamma$.  The upper panel is for the proton energies
 and the lower panel for the neutron energies.
 In each figure the left part is drawn by changing
 $\beta_2$ ($\beta_4$) from 0 to 0.42 (from 0 to 0.02)
 with keeping $\gamma=0$, while the right part by
 changing $\gamma$ from $0^\circ$ to $60^\circ$
 with keeping $(\beta_2,\beta_4)=(0.42, 0.02)$.
 The solid (dashed) lines are used for positive (negative) parity orbitals.
 }
\label{fig:multi}
\end{figure*}

In order to study the wobbling excitations of the TSD bands
in the Lu and Hf region, we have to choose suitable deformation
of the mean-field.  The standard way is to search the minima
in the Strutinsky method, but we are not yet able to perform reliable
cranked Woods-Saxon Strutinsky calculations of the potential energy surface
with fixed spin values.  The main purpose of the present work
is to confirm the existence of the wobbling excitation modes
in the microscopic Woods-Saxon RPA calculation.
Therefore, we rely on the results of
the cranked Nilsson Strutinsky calculations for choosing
the deformation parameters.
The obtained deformation parameters for the yrast TSD band of $^{163}$Lu,
i.e., the band with the parity and signature $(\pi,\alpha)=(+,1/2)$,
change very little as functions of spin;
they are about $\epsilon_2= 0.41-0.38$,
$\gamma=19^\circ-21^\circ$, and $\epsilon_4=0.03-0.05$
in the spin range $I^\pi=25/2^+- 97/2^+$.
Taking the $(\epsilon_2,\gamma,\epsilon_4)=(0.39,20^\circ,0.04)$
as typical ones, the converted values are
$(\beta_2,\gamma,\beta_4)=(0.42,18^\circ,0.02)$.
We choose these values as reference values
in the following calculations.

Now we discuss the single-particle properties of the present
Woods-Saxon potential.
In Fig.~\ref{fig:multi}
the dependences of the proton and neutron single-particle energies on
the deformation parameters $\beta_2(\beta_4)$ and $\gamma$ are shown.
It is clearly seen in the present Woods-Saxon potential
that sizable shell gaps at $Z=70\sim74$
for protons and $N=92\sim96$ for neutrons are existing
at the triaxial deformation $\gamma\approx 20^\circ$.
When cranked, these shell gaps are responsible for the TSD bands
systematically observed in nuclei in the Lu and Hf region~\cite{BenRyd04}.
Because of the relatively large deformation,
$\beta_2 > 0.4$, the $N_{\rm osc}=6$ proton orbitals resulting from
the $\pi i_{13/2}$ state come down near to the Fermi energy of
the $Z \approx 71$ system, the lowest state of which is occupied
in the yrast TSD band in Lu nuclei.
It is well-known that the occupation of
this down-sloping and highly-alignable orbitals
increase the deformation from the normal deformed,
$\beta_2 \approx 0.2$ and $\gamma\approx 0^\circ$,
to the triaxially superdeformed shape,
$\beta_2 \approx 0.4$ and $\gamma\approx 20^\circ$.

\begin{figure}[htbp]
\includegraphics[angle=0,width=80mm,keepaspectratio]{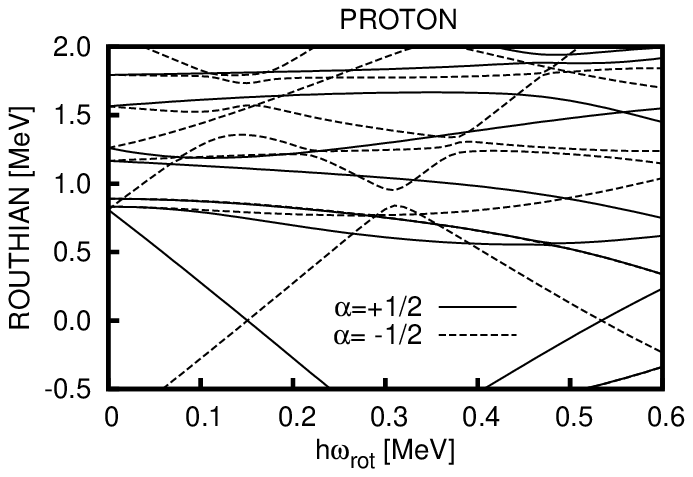}
\includegraphics[angle=0,width=80mm,keepaspectratio]{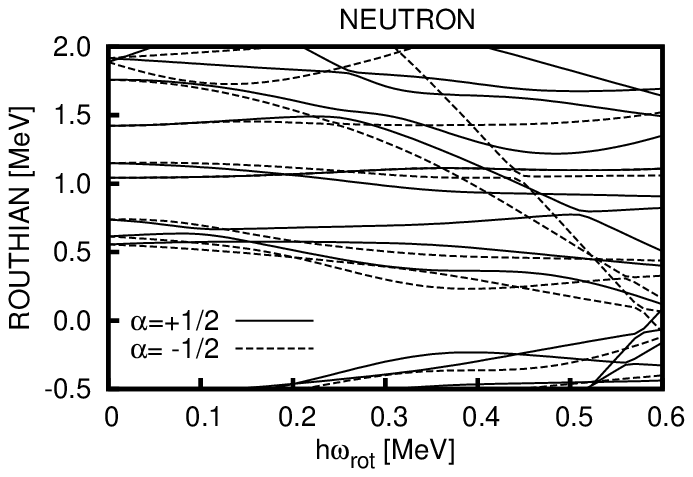}
\caption{
 The quasiparticle routhians as functions of the rotational frequency
 suitable for $^{163}$Lu.
 The fixed deformation parameters
 $(\beta_2,\gamma,\beta_4)=(0.42,18^\circ,0.02)$
 and the fixed pairing gaps ${\mit\Delta}_\pi={\mit\Delta}_\nu=0.5$ MeV
 are used.  The upper panel is for the quasiproton energies
 and the lower panel for the quasineutron energies.
 The solid (dashed) lines are used for
 signature $\alpha=1/2$ ($\alpha=-1/2$) orbitals.
 }
\label{fig:routh}
\end{figure}

The proton and neutron pairing gaps,
${\mit\Delta}_\pi$ and ${\mit\Delta}_\nu$,
respectively, are also important mean-field parameters,
for which we know very little in the TSD bands.
Therefore, we use constant values for qualitative investigation of
the TSD bands and of the wobbling excitations on them.
We will discuss the result where the pairing gaps are changed
in the following Sec.~\ref{sect:chgpair}.
The chemical potentials $\lambda_\pi$ and $\lambda_\nu$ are
always adjusted to reproduce the correct proton and neutron numbers,
$\langle N_\pi \rangle=Z$ and $\langle N_\nu \rangle=N$.
We show the quasiparticle energies (routhians) obtained
by diagonalizing the cranked mean-field hamiltonian, Eq.~(\ref{eq:crankedQP}),
in Fig.~\ref{fig:routh}, as functions of the rotational frequency.
Constant pairing gaps,
${\mit\Delta}_\pi={\mit\Delta}_\nu=0.5$ MeV,
are used with the reference deformation,
$(\beta_2,\gamma,\beta_4)=(0.42,18^\circ,0.02)$ suitable for $^{163}$Lu.
It is clear that there is no crossing of the neutron quasiparticle
before $\omega_{\rm rot}\approx 0.6$ MeV/$\hbar$.
In contrast, one highly-alignable $\alpha=1/2$ proton quasiparticle
is clearly seen;
its main component is the $N_{\rm osc}=6$ oscillator shell coming down
from the $\pi i_{13/2}$ orbitals.
This $N_{\rm osc}=6$ quasiproton is always occupied
in the lowest TSD configuration of $^{163}$Lu and
contributes to the alignment of the angular momentum about the cranking axis.
After $\omega_{\rm rot} \approx 0.6$ MeV/$\hbar$, there are many
quasiparticles, both proton and neutron ones, crossing the zero energy,
which changes the configurations of the yrast and near yrast states
and also affects the result of the RPA calculations.

In the following we will show the result of calculations mainly
for $^{163}$Lu, for which most complete experimental data are available.

\begin{figure}[htbp]
\includegraphics[angle=0,width=70mm,keepaspectratio]{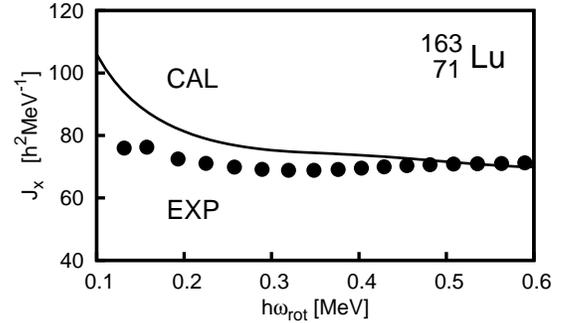}
\caption{
 The cranking moment of inertia ${\cal J}_x$ in Eq.~(\ref{eq:defJJ})
 for $^{163}$Lu as a function of the rotational frequency.
 The fixed deformation parameters
 $(\beta_2,\gamma,\beta_4)=(0.42,18^\circ,0.02)$
 and the fixed pairing gaps ${\mit\Delta}_\pi={\mit\Delta}_\nu=0.5$ MeV
 are used.
 Experimental data are taken from Ref.~\cite{wob163Lu2}.
 }
\label{fig:crJx}
\end{figure}

\begin{figure}[htbp]
\includegraphics[angle=0,width=70mm,keepaspectratio]{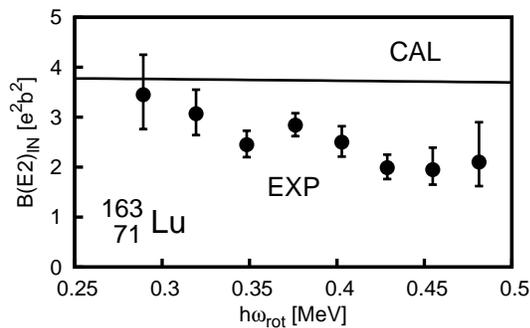}
\caption{
 The in-band $B(E2)$ calculated by Eq.~(\ref{eq:BE2in}) for $^{163}$Lu
 as a function of the rotational frequency.
 The mean-field parameters are fixed and the same as in Fig.~\ref{fig:crJx}.
 Experimental data are taken from Ref.~\cite{wob163LuQ}.
 }
\label{fig:BE2in}
\end{figure}

Next we compare the calculated and measured cranking moment of inertia
for the yrast TSD band of $^{163}$Lu in Fig.~\ref{fig:crJx},
and those of in-band $B(E2)$ in Fig.~\ref{fig:BE2in}, respectively,
as functions of the rotational frequency.
Here we use the standard procedure~\cite{BF79}
($K=1/2$ is assumed to transform the total angular momentum quantum number
to the aligned angular momentum)
for obtaining the measured cranking moment of inertia.
For the calculation in these figures
all the mean-field parameters are assumed to be constants,
$(\beta_2,\gamma,\beta_4)=(0.42,18^\circ,0.02)$ and
${\mit\Delta}_\pi={\mit\Delta}_\nu=0.5$ MeV, as a typical example.
Since the moment of inertia ${\cal J}_x$,
defined in Eq.~(\ref{eq:defJJ}),
i.e., the so-called kinematic moment of inertia,
contains the contribution of quasiproton alignment,
$\approx i/\omega_{\rm rot}$ (where $i$ is the aligned angular momentum),
the calculated ${\cal J}_x$ decreases as $\omega_{\rm rot}$ increases,
if all the mean-field parameters are fixed.
In contrast the measured inertia is almost constant against
the change of the rotational frequency,
suggesting at least one (or some) of the mean-field parameters
is actually varying.
The recently measured lifetimes of the TSD band~\cite{wob163LuQ}
indicate that the experimental in-band $B(E2)$,
on the other hand, decreases rather rapidly,
while the calculated in-band $B(E2)$ is almost constant
reflecting that the constant deformation parameters are used
in this calculation.
Therefore the $B(E2)$ data also suggest that
at least one (or some) of the deformation parameters
is changing as a function of $\omega_{\rm rot}$.
This behavior of the in-band $B(E2)$ as well as out-of-band
$B(E2)$ will be discussed in more details
in the following Sec.~\ref{sect:BE2triaxial}.

\subsection{RPA calculations}
\label{sect:calRPA}

Once the nuclear mean-field is specified and the
cranked quasiparticles, Eq.~(\ref{eq:crankedQP}), are obtained,
it is straightforward to solve the RPA equation, Eq.~(\ref{eq:EqMatF}),
or equivalently, Eq.~(\ref{eq:EqMatJ1}) for the non-NG modes.
It should be mentioned that the vacuum mean-field state
of the RPA excitations is the lowest quasiproton excited state
for the odd-$Z$ nucleus $^{163}$Lu; such a state can be
obtained by exchanging the HFB wave function and energy~\cite{RingSchuck}
of the excited quasiproton orbital to the opposite
signature sector $(U,V,E)\rightarrow (\bar{V},\bar{U},-\bar{E})$.
The details of the procedure to perform the RPA on such excited
configurations are explained in Ref.~\cite{SM83},
where the RPA excitations on the two quasineutron configuration
(the Stockholm band) were treated.
At the ground state ($\omega_{\rm rot}=0$) any one quasiparticle configuration
has the Kramers degeneracy so that it cannot be selected as a vacuum
of the RPA, but at finite rotational frequency the degeneracy is lifted
and then it is possible to perform the RPA calculation without any problems.

\begin{figure}[htbp]
\includegraphics[angle=0,width=70mm,keepaspectratio]{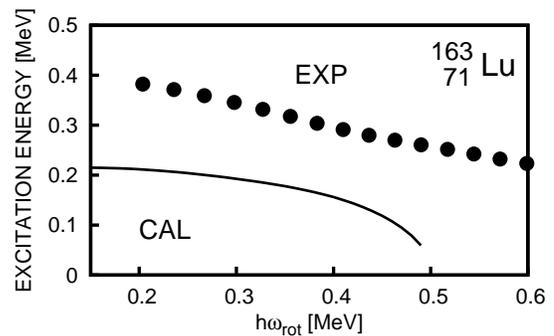}
\caption{
 The eigenenergy of the wobbling-like RPA solution
 as a function of the rotational frequency.
 The mean-field parameters are fixed and the same as in Fig.~\ref{fig:crJx}.
 Experimental data are taken from Ref.~\cite{wob163Lu2}.
 }
\label{fig:RPAom}
\end{figure}

Since it is guaranteed in our formalism that the NG mode solution
$\omega_{n={\rm NG}}=\omega_{\rm rot}$ is precisely given
by Eq.~(\ref{eq:XNG}), we actually solve Eq.~(\ref{eq:dispEqJ1}) to obtain
the non-NG eigenenergy $\omega_n$ and the amplitude
($\chi_y{\cal F}_y(n),\chi_z{\cal F}_z(n)$) by Eq.~(\ref{eq:EqMatJ1}),
and then its forward and backward amplitudes
are given by Eq.~(\ref{eq:FBamp}).
As is mentioned in the previous section, it is important to use
a sufficient model space in the RPA calculation
for the consistent description of the NG mode~(\ref{eq:XNG}).
We have checked that the full use
of the oscillator shells up to $N_{\rm max}=12$ is enough:
Then the total number of proton and neutron two quasiparticle states
$(\alpha,\beta)$ is about 210,000 for the signature ($-$) RPA calculation.
Using the formulation explained in the previous section,
we have performed the Woods-Saxon RPA calculation, and found,
just like in the previous Nilsson calculations~\cite{MSMwob1,MSMwob2,MSMwob3},
that the RPA modes that satisfy the requirements
of the wobbling motion exist systematically in the TSD configurations.

\begin{figure}[htbp]
\includegraphics[angle=0,width=70mm,keepaspectratio]{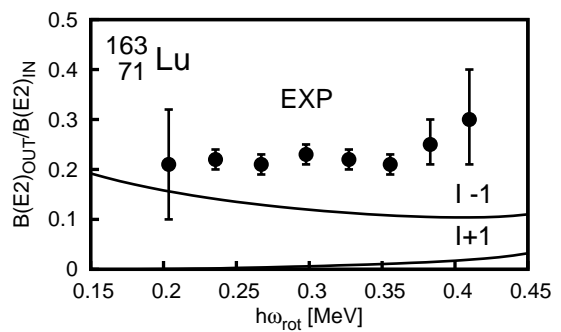}
\caption{
 The out-of-band to in-band $B(E2)$ ratio
 as a function of the rotational frequency.
 The ratios associated with the $I\rightarrow I-1$ and $I\rightarrow I+1$
 out-of-band transitions are depicted as separate lines
 with the attached labels $I-1$ and $I+1$, respectively.
 The mean-field parameters are fixed and the same as in Fig.~\ref{fig:crJx}.
 Experimental data are taken from Ref.~\cite{wob163LuQ},
 in which only the $I\rightarrow I-1$ transitions are measured.
 }
\label{fig:BE2R}
\end{figure}

\begin{figure}[htbp]
\includegraphics[angle=0,width=70mm,keepaspectratio]{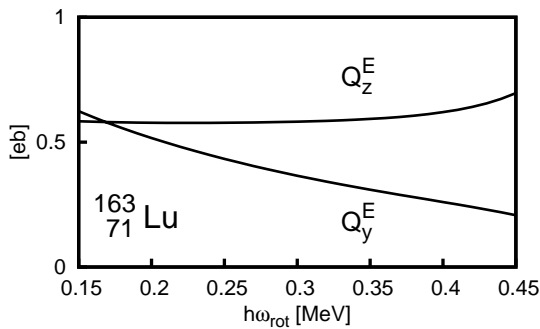}
\caption{
 The calculated quadrupole amplitudes ${\cal Q}^E_{y,z}(n={\rm wob})$
 in Eq.~(\ref{eq:BE2outQ}) of the wobbling-like RPA solution
 as functions of the rotational frequency.
 }
\label{fig:ampQyz}
\end{figure}

As an example, the excitation energy of the wobbling-like RPA
solution in $^{163}$Lu is compared with
the experimental data in Fig.~\ref{fig:RPAom}.
In this calculation the constant mean-field parameters
are used just like in Figs.~\ref{fig:crJx} and~~\ref{fig:BE2in}.
The calculated excitation energy vanishes at a finite frequency,
in this example about 0.5 MeV/$\hbar$, which is mainly due to the fact
that many quasiparticle states come close to zero energy and the vacuum
configuration changes as is seen in Fig.~\ref{fig:routh}.
The out-of-band $B(E2)$ calculated by Eqs.~(\ref{eq:BE2out})
and~(\ref{eq:E2out}) is an important quantity to identify
the wobbling motion, which is as large as about 100 Weisskopf units.
In experiments the out-of-band to in-band $B(E2)$ ratio,
$B(E2,I\rightarrow I-1)_{\rm out}/B(E2,I\rightarrow I-2)_{\rm in}$
has been directly measured, and we compare the result of RPA calculation
with experimental data for this ratio in Fig.~\ref{fig:BE2R}.

It should be noted that only the $I \rightarrow I-1$ transitions are
measured for the out-of-band transitions, which suggests that
the $I \rightarrow I+1$ out-of-band transitions are small
and indicate the positive $\gamma$ shape.
This characteristic behavior of the rotor model also exist
in the microscopic RPA calculation~\cite{SM95}.
From Eqs.~(\ref{eq:BE2out}) and~(\ref{eq:E2out})
the out-of-band $B(E2)$ of the $n$-th RPA phonon excited state is
\begin{equation}
 B(E2; In \rightarrow I\pm 1{\rm vac})\approx
  \frac{1}{2} \bigl( {\cal Q}^E_y(n) \mp {\cal Q}^E_z(n) \bigr)^2.
\label{eq:BE2outQ}
\end{equation}
The electric quadrupole matrix elements ${\cal Q}^E_y(n)$
and ${\cal Q}^E_z(n)$ have same sign and comparative magnitude
for the wobbling-like solution $(n={\rm wob})$
for the positive $\gamma$ shape,
as they are depicted in Fig.~\ref{fig:ampQyz},
so that the $I \rightarrow I-1$ transitions
are much more enhanced than the $I \rightarrow I+1$ transitions.
In Fig.~\ref{fig:BE2R} the $B(E2)$'s of both transitions are shown;
that of the $I \rightarrow I+1$ transitions is smaller
by an order of magnitude or more.

The calculated energy of the wobbling excitation is 100$-$200 keV smaller
than the experimental data;
this trend seems general and is similar
in the previous Nilsson RPA calculations~\cite{MSMwob1,MSMwob2,MSMwob3}.
It is pointed out in Refs.~\cite{HH03,TST06} that the particle-rotor
coupling increases the excitation energy of the wobbling phonon
considerably: ${\mit\Delta}\omega_{\rm p-wob} \approx j/{\cal J}_x$
in a simple limiting approximation~\cite{HH03}, where $j$ is
the angular momentum of the coupled quasiparticle
so that ${\mit\Delta}\omega_{\rm p-wob} \approx$ 90 keV
in the case of $^{163}$Lu.  In our microscopic calculation the effect
of the particle-rotor coupling is partly taken into account as a kind
of blocking effect; the vacuum of the RPA excitation contains the
odd quasiproton in the calculation.  However, there exist explicit
coupling effects between the RPA phonon and an odd quasiparticle.
The underestimation of the wobbling excitation energy in odd-$Z$
Lu nuclei in the present and the previous RPA calculations may
be due to this reason.  More detailed theoretical investigations
of the RPA-phonon-quasiparticle coupling calculation,
see e.g.~\cite{MSM88},
are necessary to draw a definite conclusion to this point.

Although the calculated out-of-band to in-band $B(E2)$ ratio
in the present Woods-Saxon RPA calculation is about 10$-$20\%
larger than the previous Nilsson RPA calculation,
it is still considerably smaller than the experimental data on average.
Moreover, in contrast to the almost constant ratio of
the experimental data, the calculated $B(E2)$ ratio is
decreasing as a function of the rotational frequency.
Thus, the basic trend is similar to the previous Nilsson calculations,
and was a most serious problem of the microscopic calculation
compared with the macroscopic rotor model.
It is, however, found that the problem of underestimation
is due to the inconsistent definition of
the triaxiality parameter ``$\gamma$''
in the mean-field potential (either Nilsson or Woods-Saxon)
and in the macroscopic model~\cite{SSMtri}.
These characteristic features of the calculated and measured $B(E2)$
will be discussed in more details
in the following Sec.~\ref{sect:BE2triaxial}.

\begin{figure}[htbp]
\includegraphics[angle=0,width=70mm,keepaspectratio]{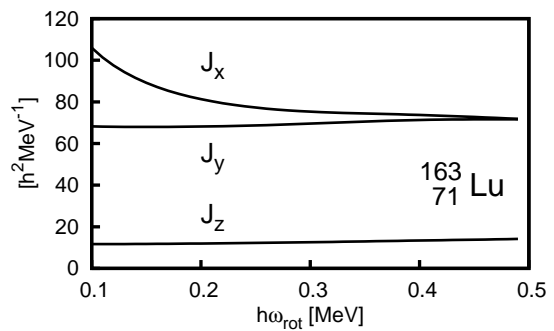}
\caption{
 The three moments of inertia, ${\cal J}_x$ and
 ${\cal J}^{\rm eff}_{y,z}(n={\rm wob})$
 as functions of the rotational frequency.
 The mean-field parameters are fixed and the same as in Fig.~\ref{fig:crJx}.
 }
\label{fig:mom3J}
\end{figure}

\begin{figure}[htbp]
\includegraphics[angle=0,width=70mm,keepaspectratio]{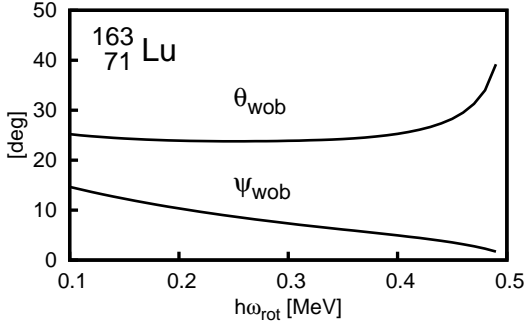}
\caption{
 The wobbling angles of the angular momentum vector in the PA frame
 for the RPA wobbling excitation,
 $\theta(n={\rm wob})$ and $\psi(n={\rm wob})$,
 defined in Eq.~(\ref{eq:wobang}),
 as functions of the rotational frequency.
 }
\label{fig:wobangle}
\end{figure}

Now let us discuss the results of calculation transformed into the PA frame.
The three moments of inertia corresponding
to the wobbling-like solution in Fig.~\ref{fig:RPAom} are shown
in Fig.~\ref{fig:mom3J}.
The ${\cal J}_x$ inertia is the same as in Fig.~\ref{fig:crJx}.
Other inertias, ${\cal J}^{\rm eff}_y$ and ${\cal J}^{\rm eff}_z$, are
calculated according to Eq.(\ref{eq:effJ}) not to Eq.(\ref{eq:effJS});
the results of these two equations
are different for the general mean-field potential.
We have found, however, that the actual difference between the
two definitions is small, within 1\%
at least in the present calculation; this suggests that
the operators $F_{y,z}$ in Eq.~(\ref{eq:defF}) obtained with
the present Woods-Saxon mean-field are not so much different
from those proportional to the quadrupole operators $Q_{y,z}$
in Eq.~(\ref{eq:defQ}).
The calculated inertias ${\cal J}^{\rm eff}_{y,z}$ are almost constant
in this calculation with fixed mean-field parameters,
and ${\cal J}^{\rm eff}_y$ is much larger than ${\cal J}^{\rm eff}_z$.
These basic features are similar to
the previous Nilsson RPA calculations~\cite{MSMwob1}.
We also show, in Fig.~\ref{fig:wobangle},
the wobbling angles of the angular momentum vector in the PA frame
(c.f., Eqs.(\ref{eq:AngQ}),~(\ref{eq:rQPA}) and~(\ref{eq:rJPA})),
which are defined for the $n$-th RPA mode by~\cite{MO04,Mat08},
\begin{equation}
 \begin{array}{ll}
 \tan\theta(n)
 &\equiv
 {\displaystyle
 \frac{\sqrt{(J_y)_{\rm PA}(n)^2+(J_z)_{\rm PA}(n)^2}}
    {\langle J_x \rangle} }\\
 &= \sqrt{r^2_y(n)+r^2_z(n)}, \vspace*{2mm} \\
 \tan\psi(n)
 &\equiv
 {\displaystyle
 \frac{(J_z)_{\rm PA}(n)}{(J_y)_{\rm PA}(n)}
 =\frac{r_y(n)}{r_z(n)}},
 \end{array}
\label{eq:wobang}
\end{equation}
for which the small amplitude approximation requires
$\tan\theta(n) \approx \theta(n)$.
From this figure the angle $\theta$ is about $25^\circ$ (0.44 radian),
and the small amplitude approximation may be acceptable.
Note that the quantities $r_y(n)$ and $r_z(n)$ describe
the shape fluctuations, Eq.~(\ref{eq:rQPA}), and, at the same time,
the angular momentum fluctuations, Eq.~(\ref{eq:rJPA});
$r_y(n) < r_z(n)$ in this calculation and $r_y(n)$ decreases rapidly
as a function of $\omega_{\rm rot}$ (c.f., Fig.~\ref{fig:ampQyz}),
and $r_y(n) \ll r_z(n)$ at higher frequency.
This means that the angular momentum fluctuation
of the $y$-direction is much larger than that of the $z$-direction,
and so the angle $\psi \rightarrow 0$ in Fig.~\ref{fig:wobangle}.

\begin{figure}[htbp]
\includegraphics[angle=0,width=70mm,keepaspectratio]{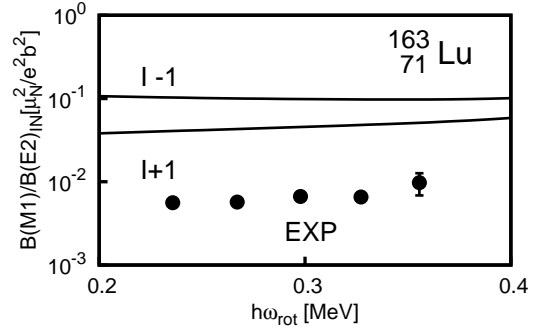}
\caption{
 The out-of-band $B(M1)$ to in-band $B(E2)$ ratio
 as a function of the rotational frequency.
 Both the $I\rightarrow I-1$ and $I\rightarrow I+1$ transitions
 are depicted as separate lines with the attached labels $I-1$ and $I+1$,
 respectively.
 Note that the ratios are drawn in a log scale.
 The mean-field parameters are fixed and the same as in Fig.~\ref{fig:crJx}.
 Experimental data are taken from Ref.~\cite{Hamamoto01},
 in which only the $I\rightarrow I-1$ transitions are measured.
 }
\label{fig:BM1R}
\end{figure}

Finally, the calculated out-of-band $B(M1)$ to in-band $B(E2)$ ratio
is compared with the experimental data
in Fig.~\ref{fig:BM1R} as in the same way as the $B(E2)$ ratio,
but in the log scale.  The $B(M1)$ transition is
calculated by Eqs.~(\ref{eq:BM1out}) and~(\ref{eq:M1out}), i.e.,
\begin{equation}
 B(M1; In \rightarrow I\pm 1{\rm vac})\approx
  \frac{1}{2} \bigl( \mu_y(n) \pm \mu_z(n) \bigr)^2,
\label{eq:BE2outMU}
\end{equation}
where the RPA matrix elements of the $M1$ operator are given
in the same way as those of $E2$ operator,
\begin{equation}
 \mu_y(n)\equiv \langle [X_n,i\mu_y] \rangle,\quad
 \mu_z(n)\equiv \langle [X_n,\mu_z] \rangle.
\label{eq:MUamp}
\end{equation}
The effective spin $g$-factor $g^{\rm eff}_s = 0.6\,g^{\rm free}_s$
is used~\cite{wob163Lu2,Ham02};
we have checked that the result with using
$g^{\rm eff}_s = 0.7\,g^{\rm free}_s$ does not change essentially.
The $I\rightarrow I-1$ transitions are larger than
the $I\rightarrow I+1$ transitions by factor two to four,
in contrast to the case of $B(E2)$ ratio.
This is because the $\mu_y(n)$ amplitude is much larger than
the $\mu_z(n)$ at all rotational frequencies in the present calculation.
As shown in Fig.~\ref{fig:BM1R}, the calculated ratio
is about one order of magnitude larger than the experimental data,
which is similar to that of Ref.~\cite{AND06}.
The relative sign of the matrix elements between the $E2$ and $M1$
transitions is negative, which agrees with the experimentally
measured $E2/M1$ mixing ratio~\cite{wob163Lu1,wob163Lu2}.
As for the $B(M1)$ ratio, the measured values~\cite{Hamamoto01}
are 0.0056 [$\mu_N^2/e^2b^2$] at lower spins and 0.0098 at higher spins.
The particle-rotor model calculations~\cite{wob163Lu1,TST06}
give about 0.015$-$0.02 at lower spins and about 0.01 at higher spins.
In the previous Nilsson RPA calculation~\cite{MSMwob1},
the corresponding values are about 0.04 and 0.015.
Compared with these other calculations, our present value of $B(M1)$ is
too large, about 0.1, and almost constant,
while the other calculations predict that the $B(M1)$ ratio
decreases as a function of $\omega_{\rm rot}$.
We do not understand the reason why the present Woods-Saxon RPA
calculation gives larger values especially at higher spin than
the previous Nilsson RPA calculation.
It should be mentioned that the absolute value of
the $B(M1)$ is small compared to the Weisskopf unit,
but still it is much larger than the experimental data.
This means that the RPA eigenmode operator
corresponding to the wobbling excitation should be almost
completely orthogonal to the $M1$ operator~(\ref{eq:M1out}),
i.e., the various orbital contributions should cancel out,
in order to obtain the agreement with the experimental data;
the cancellation is not enough in the present calculation.
Since our calculation does not agree with experimental data
for $B(M1)$, we concentrate on the excitation energy and $B(E2)$
in the following further investigations.

\subsection{Dependence on mean-field parameters}
\label{sect:depmeanfield}

Until now we have shown the results of the microscopic calculations
with fixed mean-field parameters at the reference values,
$(\beta_2,\gamma,\beta_4)=(0.42,18^\circ,0.02)$ and
${\mit\Delta}_\pi={\mit\Delta}_\nu=0.5$ MeV.
As it is explained in Sec.~\ref{sect:calMF}, we do not know
very well about what are the best values for these parameters.
They might change depending on the rotational frequency,
and then it must be clarified how their dependences are;
for example, it is well-known that the proton and neutron pairing gaps
should decrease as functions of the rotational frequency
because of the Coriolis anti-pairing effect.
Instead of looking for what should be the values of these
mean-field parameters,
we here discuss how the results of the present Woods-Saxon RPA
calculation change when these mean-field parameters are varied
in a reasonable range.

\begin{figure}[htbp]
\includegraphics[angle=0,width=70mm,keepaspectratio]{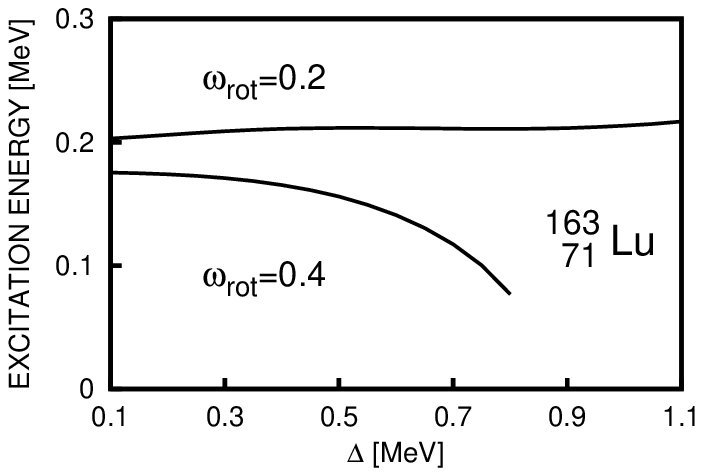}
\includegraphics[angle=0,width=70mm,keepaspectratio]{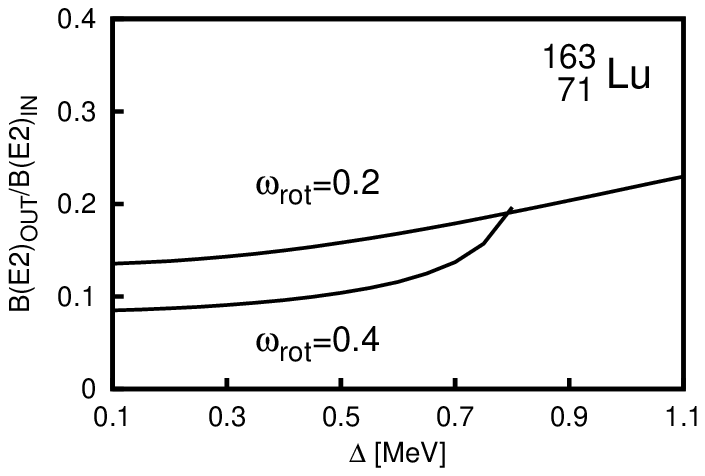}
\caption{
 Dependence of
 the eigenenergy (upper panel) or of
 the out-of-band to in-band $B(E2)$ ratio (lower panel)
 on the pairing gap ${\mit\Delta}={\mit\Delta}_\pi={\mit\Delta}_\nu$
 for the wobbling-like RPA solution.
 The two curves correspond to the results
 at $\omega_{\rm rot}=0.2$ and 0.4 MeV/$\hbar$.
 The mean-field parameters except the pairing gaps
 are the same as in Fig.~\ref{fig:crJx}.
 }
\label{fig:RPAdepdel}
\end{figure}

First, we show the dependence of the wobbling excitation energy
and the $B(E2)$ ratio on the pairing gap,
${\mit\Delta}\equiv{\mit\Delta}_\pi={\mit\Delta}_\nu$,
i.e., the proton and neutron gaps being set equal for simplicity,
in Fig.~\ref{fig:RPAdepdel}.
In these figures, those at the rotational frequencies,
$\omega_{\rm rot}=0.2$ and 0.4 MeV/$\hbar$ are shown.
As it is clear the dependence of the wobbling excitation energy
on the pairing gap is rather weak, and the very collective RPA
solution exists even in the ${\mit\Delta}=0$ limit;
at higher frequency $\omega_{\rm rot}=0.4$ MeV/$\hbar$ near
the critical frequency of vanishing RPA energy,
$\omega_{\rm rot} \approx 0.5$ MeV/$\hbar$,
the solution becomes unstable for larger pairing gaps.
This is in contrast to the collective vibrational mode,
and discussed~\cite{MSMwob1,MSMwob3,SMMprec}
to be the characteristic property of the wobbling mode,
which is of rotational origin.
The dependence of the $B(E2)$ ratio on $\Delta$, which comes solely from
the out-of-band $B(E2)$, is slightly stronger, but again,
smaller than that of the collective vibrational mode.
The dependence of the cranking moment of inertia and
the in-band $B(E2)$ are well-known:
The cranking moment of inertia decreases
by about 27\% (15\%) at $\omega_{\rm rot}=0.2$ (0.4) MeV/$\hbar$
as ${\mit\Delta}$ increases from 0.1 to 0.8 MeV
for $^{163}$Lu (not shown),
while the in-band $B(E2)$
does not depend on ${\mit\Delta}$ essentially.

\begin{figure}[htbp]
\includegraphics[angle=0,width=70mm,keepaspectratio]{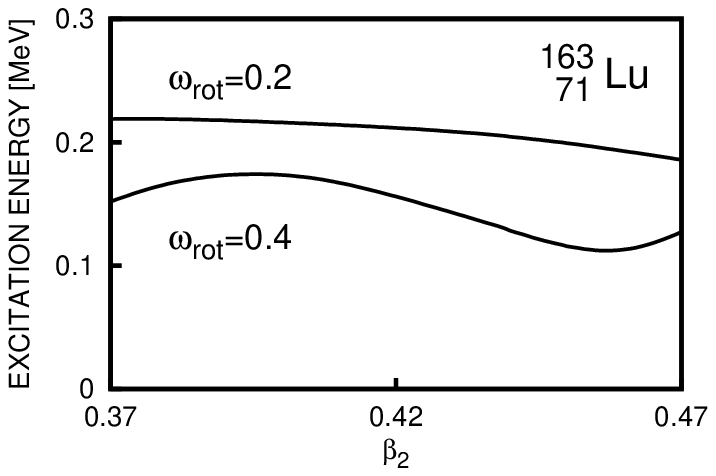}
\includegraphics[angle=0,width=70mm,keepaspectratio]{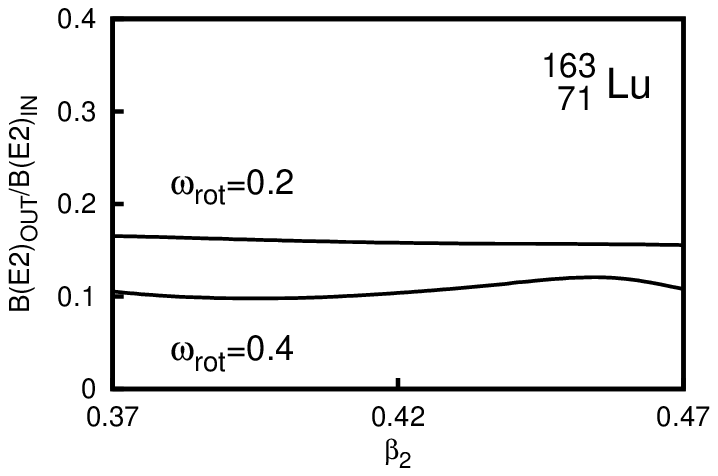}
\caption{
 Dependence of
 the eigenenergy (upper panel) or of
 the out-of-band to in-band $B(E2)$ ratio (lower panel)
 on the deformation parameter $\beta_2$
 for the wobbling-like RPA solution.
 The two curves correspond to the results
 at $\omega_{\rm rot}=0.2$ and 0.4 MeV/$\hbar$.
 The mean-field parameters except $\beta_2$
 are the same as in Fig.~\ref{fig:crJx}.
 }
\label{fig:RPAdepb2}
\end{figure}

Next, we show the dependence of the wobbling excitation energy
and the $B(E2)$ ratio on the magnitude of deformation $\beta_2$
in Fig.~\ref{fig:RPAdepb2}.
Other parameters are fixed at the reference values.
The excitation energy slightly depends on $\beta_2$ but
is stable in the calculated range.
The $B(E2)$ ratio is almost independent of $\beta_2$;
this is quite reasonable in the macroscopic model,
because both the in-band $B(E2)$ and out-of-band $B(E2)$ depend
on the square of magnitude of deformation and it cancels
when their ratio is taken.
The in-band $B(E2)$ increases monotonically
as $\beta_2$ increases with keeping $\gamma$ (not shown),
because the intrinsic quadrupole moment about the $x$-axis increases.
However, the cranking moment of inertia ${\cal J}_x$ is almost constant or
slightly decreases (not shown); this is because the core contribution
increases but the quasiproton alignment contribution decreases
and these two almost cancel.

\begin{figure}[htbp]
\includegraphics[angle=0,width=70mm,keepaspectratio]{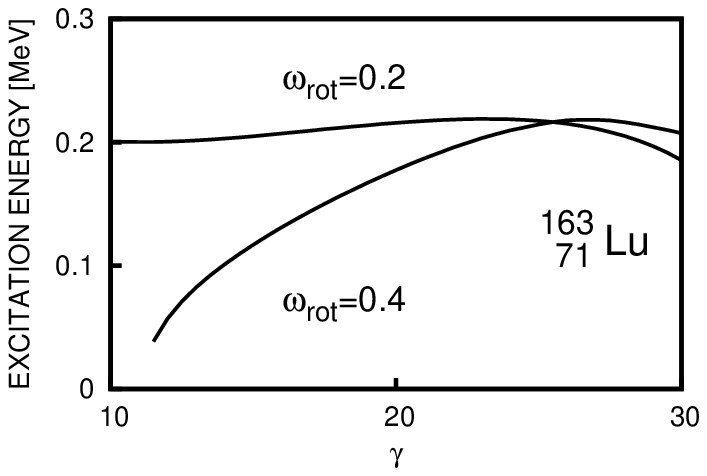}
\includegraphics[angle=0,width=70mm,keepaspectratio]{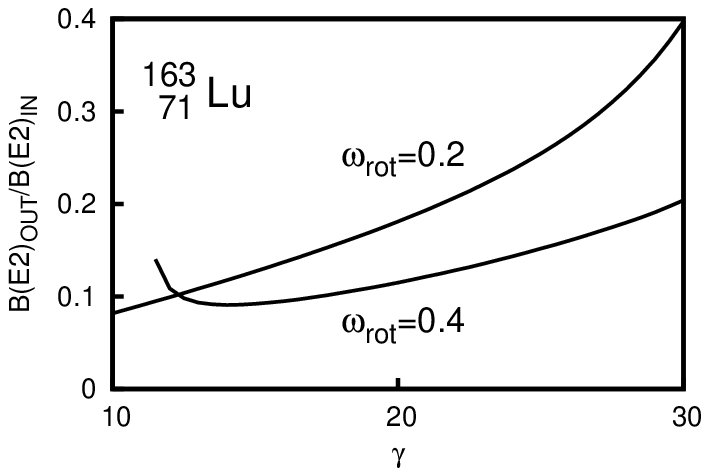}
\caption{
 Dependence of
 the eigenenergy (upper panel) or of
 the out-of-band to in-band $B(E2)$ ratio (lower panel)
 on the deformation parameter $\gamma$
 for the wobbling-like RPA solution.
 The two curves correspond to the results
 at $\omega_{\rm rot}=0.2$ and 0.4 MeV/$\hbar$.
 The mean-field parameters except $\gamma$
 are the same as in Fig.~\ref{fig:crJx}.
 }
\label{fig:RPAdepgam}
\end{figure}

Finally, we show the dependence of the wobbling excitation energy
and the $B(E2)$ ratio on the triaxiality parameter $\gamma$
in Fig.~\ref{fig:RPAdepgam}.
Again, the excitation energy slightly depends on $\gamma$
but is stable in the range, $15^\circ < \gamma < 30^\circ$.
On the other hand, the $B(E2)$ ratio increases rapidly
as $\gamma$ increases.
Both the in-band $B(E2)$ and out-of-band $B(E2)$ depend on
the triaxiality in definite ways according to the macroscopic model.
The result of the microscopic RPA calculation reproduces
the characteristic features of the rotor model rather well,
and this is precisely the reason that we can identify the obtained
RPA solution as the wobbling excitation mode.
The in-band $B(E2)$ decreases monotonically as $\gamma$ increases
(not shown, but see Sec.~\ref{sect:BE2triaxial}).
The cranking moment of inertia slightly increases (not shown);
the core contribution decreases but the quasiproton alignment contribution
increases, and the alignment contribution slightly wins in the present case.

\subsection{Calculation with varying pairing gaps}
\label{sect:chgpair}

The mean-field parameters are kept constants against the change
of the rotational frequency in the results presented above.
However, at least the proton and neutron pairing gaps should decrease
as functions of the rotational frequency due to the Coriolis
anti-pairing effect.
Therefore, it is desirable to determine
${\mit\Delta}_\pi$ and ${\mit\Delta}_\nu$ selfconsistently.
For this purpose, we fix the pairing force strength at
the ground state with the pairing model space
explained in Sec.~\ref{sect:calMF}, by
\begin{equation}
 G_\tau={\mit\Delta}_\tau({\rm eo}) / \langle P_\tau \rangle_{\rm gr},
\label{eq:pairG}
\end{equation}
and use this strength for the TSD bands.
Here ${\mit\Delta}_\tau({\rm eo})$ is the experimental
even-odd mass difference
evaluated by the third order difference formula~\cite{BMtext75},
and $\langle P_\tau \rangle_{{\rm gr}}$ is the expectation value
of the monopole pair transfer operator with respect to the normal
deformed ground state ($\omega_{\rm rot}=0$),
whose deformation is obtained by the Woods-Saxon Strutinsky calculation;
i.e., $\beta_2=0.217,\gamma=0,\beta_4=-0.002$ for $^{163}$Lu.

The same Strutinsky calculation for the ground state,
where the smooth pairing gap method is used with
$\widetilde{\mit\Delta}=13/\sqrt{A}$ MeV, gives
the pairing gaps, ${\mit\Delta}_{\pi,\rm gr}=0.839$ MeV
and ${\mit\Delta}_{\nu,\rm gr}=1.233$ MeV for $^{163}$Lu.
It should be noted, here, that the situation is puzzling:
the even-odd mass differences for $^{163}$Lu are
${\mit\Delta}_{\pi}({\rm eo})=1.243$ MeV
and ${\mit\Delta}_{\nu}({\rm eo})=0.868$ MeV
by using the 1995 mass data~\cite{AW95}.
In the neighboring even-even nuclei,
${\mit\Delta}_{\pi}({\rm eo})=1.180$ MeV
and ${\mit\Delta}_{\nu}({\rm eo})=1.167$ MeV for $^{162}$Yb,
and ${\mit\Delta}_{\pi}({\rm eo})=1.260$ MeV
and ${\mit\Delta}_{\nu}({\rm eo})=1.221$ MeV for $^{164}$Hf.
Taking into account the blocking effect,
the calculated pairing gaps (${\mit\Delta}_{\pi,\rm gr}=0.839$ MeV
and ${\mit\Delta}_{\nu,\rm gr}=1.233$ MeV)
are reasonable in comparison with
the even-odd mass differences of neighboring even-even nuclei.
However, ${\mit\Delta}_{\pi}({\rm eo})$ in $^{163}$Lu is too large
while ${\mit\Delta}_{\nu}({\rm eo})$ is too small compared with
those in the neighboring even-even nuclei.
This trend is not particular in $^{163}$Lu, but it seems general
in odd-proton nuclei in this rare earth region,
for which we do not know the reason yet.
In Ref.~\cite{SS06}, we have used the neutron or proton pairing force strength
which is an average of those for the neighboring even-even nuclei
determined by the even-odd mass differences.
In order to confirm that the results are not very different
from those of Ref.~\cite{SS06}, we use, in the present work,
the pairing force strengths determined by
the even-odd mass differences of $^{163}$Lu.

\begin{figure}[htbp]
\includegraphics[angle=0,width=70mm,keepaspectratio]{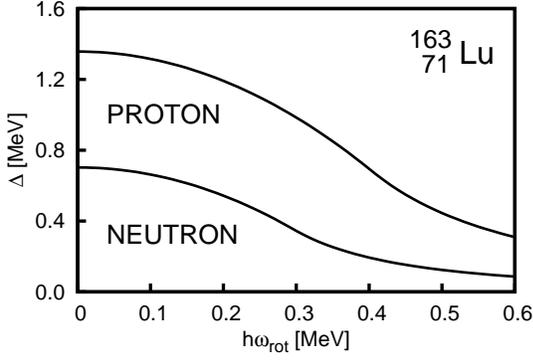}
\caption{
 The proton and neutron pairing gaps used in the pairing varied calculation
 as functions of the rotational frequency for $^{163}$Lu.
 The deformation parameters used are
 $(\beta_2,\gamma,\beta_4)=(0.42,18^\circ,0.02)$.
 }
\label{fig:gaprot}
\end{figure}

By using the pairing force strengths thus determined,
we have done selfconsistent pairing calculation with
the deformation parameters fixed at the reference value
$(\beta_2,\gamma,\beta_4)=(0.42,18^\circ,0.02)$.
Then the proton and neutron pairing gaps vanish at
$\omega_{\rm rot}\approx 0.42$ and 0.43 MeV/$\hbar$, respectively,
due to the alignments of quasiparticles.
The abrupt pairing collapse is a drawback of the mean-field approximation,
and we use the following phenomenological parameterization
at finite rotational frequency~\cite{WSNJ90},
\begin{equation}
 {\mit\Delta}(\omega_{\rm rot})={\mit\Delta}_0\times\left\{\begin{array}{ll}
 \left[1-
 {\displaystyle \frac{1}{2}\left(
 \frac{\omega_{\rm rot}}{\omega_{\rm c}}\right)^2}\right],
 & \quad\omega_{\rm rot} < \omega_{\rm c}, \\
 \,\,\,{\displaystyle \frac{1}{2}\left(
 \frac{\omega_{\rm c}}{\omega_{\rm rot}}\right)^2},
 & \quad\omega_{\rm rot} \ge \omega_{\rm c},
\end{array}\right.
\label{eq:Delpar}
\end{equation}
where the parameters ${\mit\Delta}_0={\mit\Delta}(\omega_{\rm rot}=0)$
and $\omega_c$, such that
${\mit\Delta}(\omega_{\rm rot}=\omega_c)=\frac{1}{2}{\mit\Delta}_0$,
are determined by the selfconsistent pairing calculations.
The resultant pairing gaps are shown in Fig.~\ref{fig:gaprot}.

\begin{figure}[htbp]
\includegraphics[angle=0,width=70mm,keepaspectratio]{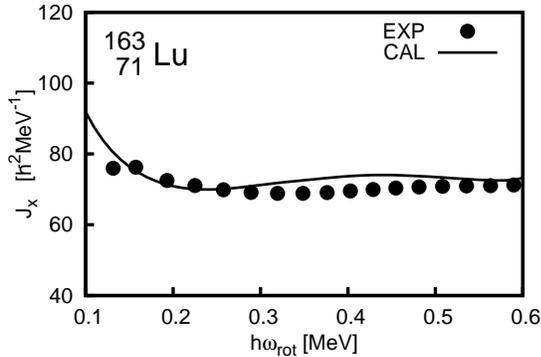}
\caption{
 The cranking moment of inertia as a function of the rotational frequency.
 The deformation parameters are fixed but the pairing gaps in
 Fig.~\ref{fig:gaprot} are used.
 }
\label{fig:crJxdel}
\end{figure}

We have done the semi-phenomenological pairing selfconsistent
calculation by using the pairing gaps in Fig.~\ref{fig:gaprot}
and the constant deformation parameters at the reference values,
$(\beta_2,\gamma,\beta_4)=(0.42,18^\circ,0.02)$.
The result for the cranking moment of inertia
is shown in Fig.~\ref{fig:crJxdel}.
Comparing with the constant pairing calculation in Fig.~\ref{fig:crJx},
the agreement between the experimental data and the calculation
is much better.  This means that the present Woods-Saxon RPA model
can describe both the yrast TSD band and the wobbling excitation
at the same time, which is a great advantage over
the previous Nilsson RPA model, where the cranking moment of inertia
is overestimated and the Strutinsky renormalization of the angular
momentum is necessary to reproduce the yrast TSD band.

\begin{figure}[htbp]
\includegraphics[angle=0,width=70mm,keepaspectratio]{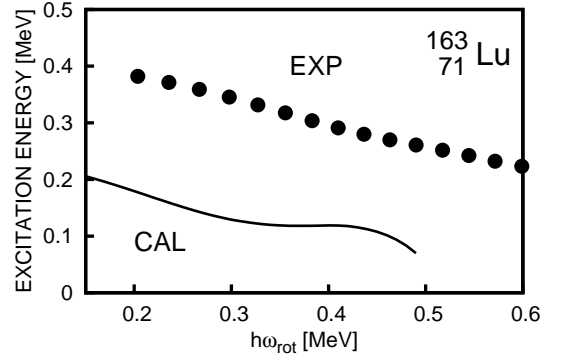}
\caption{
 The eigenenergy of the wobbling-like solution
 as a function of the rotational frequency.
 The deformation parameters are fixed but the pairing gaps in
 Fig.~\ref{fig:gaprot} are used.
 }
\label{fig:RPAomdel}
\end{figure}

\begin{figure}[htbp]
\includegraphics[angle=0,width=70mm,keepaspectratio]{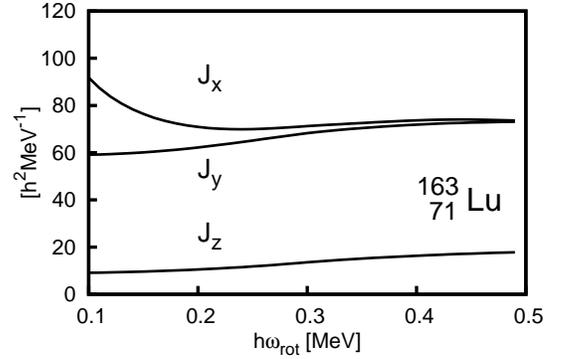}
\caption{
 The three moments of inertia, ${\cal J}_x$ and
 ${\cal J}^{\rm eff}_{y,z}(n={\rm wob})$
 as functions of the rotational frequency.
 The deformation parameters are fixed but the pairing gaps in
 Fig.~\ref{fig:gaprot} are used.
 }
\label{fig:mom3Jdel}
\end{figure}

\begin{figure}[htbp]
\includegraphics[angle=0,width=70mm,keepaspectratio]{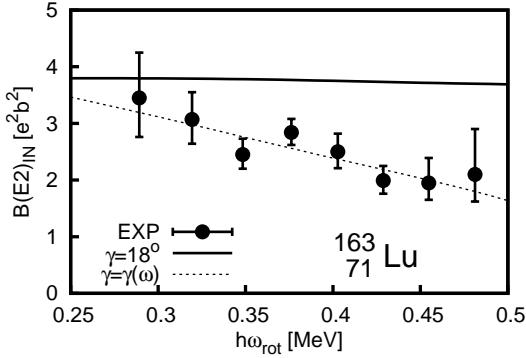}
\caption{
 The in-band $B(E2)$ defined for $^{163}$Lu
 as a function of the rotational frequency (solid line).
 The deformation parameters are fixed but the pairing gaps in
 Fig.~\ref{fig:gaprot} are used.
 The dashed line is the result of calculation with further changing
 the $\gamma$ parameter linearly from $\gamma=18^\circ$
 at $\omega_{\rm rot}=0.2$ MeV/$\hbar$ to
 $\gamma=30^\circ$ at $\omega_{\rm rot}=0.4$ MeV/$\hbar$.
 }
\label{fig:BE2indel}
\end{figure}

\begin{figure}[htbp]
\includegraphics[angle=0,width=70mm,keepaspectratio]{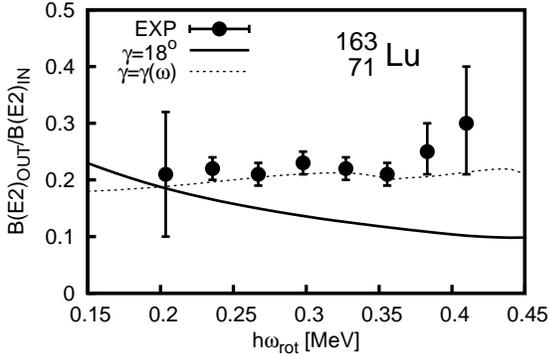}
\caption{
 The out-of-band to in-band $B(E2)$ ratio for $^{163}$Lu
 as a function of the rotational frequency (solid line).
 The deformation parameters are fixed but the pairing gaps in
 Fig.~\ref{fig:gaprot} are used.
 The dashed line is the result of calculation with further changing
 the $\gamma$ parameter linearly from $\gamma=18^\circ$
 at $\omega_{\rm rot}=0.2$ MeV/$\hbar$ to
 $\gamma=30^\circ$ at $\omega_{\rm rot}=0.4$ MeV/$\hbar$.
 The $B(E2)$ ratio with only $I \rightarrow I-1$ transitions
 is shown.
 }
\label{fig:BE2Rdel}
\end{figure}

The result of the excitation energy of the RPA wobbling mode
is compared with the experimental data in Fig.~\ref{fig:RPAomdel}.
The behavior of the excitation energy is not very different from
the constant pairing calculation in Fig.~\ref{fig:RPAom},
but the overall energy is slightly lower
and about 200 keV smaller than the experimental data.
The three moments of inertia in the PA frame calculated
with varying pairing gaps are shown in Fig.~\ref{fig:mom3Jdel}.
In contrast to the corresponding constant pairing calculation
in Fig.~\ref{fig:mom3J}, the inertias ${\cal J}^{\rm eff}_{y,z}$
acquire sizable dependence on the rotational frequency,
but the basic features such as
${\cal J}^{\rm eff}_y \gg {\cal J}^{\rm eff}_z$ are almost the same.
As for the $B(E2)$'s, the in-band $B(E2)$ and the out-of-band to in-band
$B(E2)$ ratio are shown as solid lines
in Figs.~\ref{fig:BE2indel} and~\ref{fig:BE2Rdel}.
As it is discussed in Sec.~\ref{sect:depmeanfield},
the in-band $B(E2)$ does not depend on the pairing gaps,
so the result is almost the same as that with the fixed mean-field
parameters in Fig.~\ref{fig:BE2in}.
The $B(E2)$ ratio is slightly different from that in Fig.~\ref{fig:BE2R},
but the difference is small.
Thus, the calculation with varying pairing gaps is not helpful
to fill the gap for the discrepancy between the calculation and
the experimental data of $B(E2)$'s, which is considered in
the next subsection.

\subsection{\mbox{\boldmath{$B(E2)$}} and triaxial deformation}
\label{sect:BE2triaxial}

The out-of-band $B(E2)$ from the excited band to the yrast TSD band
is the crucial quantity to identify the nuclear wobbling motion.
The recent lifetime measurement of the excited and
yrast TSD bands~\cite{wobLuQ,wob163LuQ} revealed that
the out-of-band $B(E2)$ is as large as about 100 Weisskopf units,
which is much larger than the typical collective-vibrational $E2$ transitions,
and is possibly the largest interband $E2$ transitions.
It is important not only to identify the wobbling excitation
but also to deduce the triaxiality parameter $\gamma$;
according to the rotor model, the out-of-band to in-band $B(E2)$ ratio
directly reflects the triaxiality of the quadrupole deformation.
Here it should be noted that there are various different definitions
of the triaxiality parameter $\gamma$, and one should be careful
which definition is used for quantitative discussions~\cite{SSMtri}.

The parameter $\gamma$ in the mean-field potential,
Eq.~(\ref{eq:WSbeta}), is one of such definitions,
called $\gamma(\mbox{pot:WS})$ in Ref.~\cite{SSMtri},
but it is different from the one specified by the quadrupole moments,
which is called $\gamma(\mbox{den})$ and defined by
\begin{equation}
\left\{\begin{array}{l}
 Q\cos\gamma\equiv\sqrt{\frac{16\pi}{5}}\,
 \langle Q^{(+)}_{20} \rangle,\\
 Q\sin\gamma\equiv-\sqrt{\frac{16\pi}{5}}\,
 \langle Q^{(+)}_{22} \rangle,
 \end{array}\right.
\label{eq:Qintr}
\end{equation}
where $Q$ is the total quadrupole moment.
In terms of this $\gamma=\gamma(\mbox{den})$, the quantities
$\alpha_{y,z} R^2$ in Eq.~(\ref{eq:defP}) can be expressed
as $\alpha_y R^2 = -\sqrt{\frac{5}{16\pi}}\,Q\sin(\gamma+60^\circ)$
and $\alpha_z R^2 = -\sqrt{\frac{5}{16\pi}}\,Q\sin\gamma$,
and then by using Eq.~(\ref{eq:ampr3J}) with~(\ref{eq:rQPA})
the out-of-band $B(E2)$ in Eq.~(\ref{eq:BE2outQ}) is represented as
in the same form as the macroscopic rotor expression~\cite{BMtext75},
\begin{equation}
 \begin{array}{l}
 B(E2; In \rightarrow I\pm 1{\rm vac})\approx
 {\displaystyle
  \frac{5}{16\pi}\left(e\frac{Z}{A}Q\right)^2\frac{1}{I}
  }\\
 \times
 {\displaystyle
 \left[\left(\frac{W_z(n)}{W_y(n)}\right)^{1/4}\sin(\gamma+60^\circ)
 \mp\left(\frac{W_y(n)}{W_z(n)}\right)^{1/4}\sin\gamma \right]^2,
  } \end{array}
\label{eq:BE2outRotor}
\end{equation}
with the definitions
\begin{equation}
 W_y(n)\equiv 1/{\cal J}^{\rm eff}_z(n) - 1/{\cal J}_x,\quad
 W_z(n)\equiv 1/{\cal J}^{\rm eff}_y(n) - 1/{\cal J}_x.
\label{eq:defW}
\end{equation}
Here $\langle J_x \rangle \approx I$, $Q^E_{y,z}=(eZ/A)\,Q_{y,z}$,
and $c_n^2=1$ as well as $\sigma=+$ in Eq.(\ref{eq:sigmac})
are further assumed.
As it is discussed in Sec.~\ref{sect:IntpPA},
the condition $c_n^2=1$ and $\sigma=+$ can be regarded as
a requirement for the wobbling-like RPA solution: In the present
and the previous calculations this condition is satisfied
if the used model space is large enough to guarantee
the exact NG mode decoupling;
the deviation is within 2\% in the present calculations.
Using the same $\gamma=\gamma(\mbox{den})$ and the assumption
$Q^E_{2\nu}=(eZ/A)Q_{2\nu}$, the in-band $B(E2)$ in Eq.(\ref{eq:BE2in})
can be also expressed as
\begin{equation}
 B(E2; I \rightarrow I\pm 2)\approx
  \frac{5}{32\pi}\left(e\frac{Z}{A}Q\right)^2\,\cos^2(\gamma+30^\circ).
\label{eq:BE2inRotor}
\end{equation}
In this way, the out-of-band to in-band $B(E2)$ ratio does not depend
on the total quadrupole moment $Q$ (and so not on $\beta_2$ and $\beta_4$)
in a good approximation, and the decreasing trend of
the calculated $B(E2)$ ratio
as a function of the rotational frequency can be naturally understood
by the $1/I$ factor in Eq.~(\ref{eq:BE2outRotor}).

It should be emphasized that the $\gamma$ parameter that appears in
Eqs.~(\ref{eq:BE2outRotor}) and~(\ref{eq:BE2inRotor}),
which are essentially the formula employed in the particle-rotor
model calculations~\cite{Ham02,HH03,TST06,TST08}
(note that the effect of the odd particle is negligible for $B(E2)$'s),
is $\gamma=\gamma(\mbox{den})$ in Eq.~(\ref{eq:Qintr})
and not $\gamma(\mbox{pot:WS})$ in Eq.~(\ref{eq:WSbeta}).
As it is demonstrated in Ref.~\cite{SSMtri},
they are rather different at a given shape especially
for larger quadrupole deformations;
for example, $\gamma(\mbox{den})=20^\circ$ corresponds to
$\gamma(\mbox{pot:WS})\approx 30^\circ$,
for the TSD shape of the Lu and Hf region (see Fig.~4 of Ref.~\cite{SSMtri}).
It was discussed in the particle-rotor model calculations~\cite{Ham02,HH03}
that $\gamma(\mbox{den})\approx 20^\circ$ are necessary
in order to reproduce the experimental $B(E2)$ ratio on average.
This means that one has to use $\gamma\approx 30^\circ$ in the
Woods-Saxon potential in Eq.(\ref{eq:WSbeta}).
Since the calculated $B(E2)$ ratio increases as a function of
the parameter $\gamma$ as is discussed in Sec.~\ref{sect:depmeanfield},
we obtain similar amount of agreement to that in the particle-rotor model
if the increased value $\gamma\approx 30^\circ$ is used:
This is demonstrated for the Nilsson RPA calculation in Ref.~\cite{SSMtri}.

Although the average magnitude is improved by increasing
the triaxiality of the mean-field, there is still
considerable disagreement between the calculated and measured $B(E2)$ ratios:
Their rotational frequency dependences are different.
This suggest that some parameter of the mean-field should change
as a function of the rotational frequency.
Here it should be mentioned that the measured in-band $B(E2)$
also depend on the rotational frequency;
it decreases as $\omega_{\rm rot}$ increases.
This trend can be understood either as a result of decreasing
$\beta_2$ ($\beta_4$) or as a result of increasing $\gamma$ (or both of them).
However, the $B(E2)$ ratio is almost independent on $\beta_2$ ($\beta_4$).
Therefore, a simplest way to understand both $B(E2)$ data is
to increase $\gamma$ as a function of $\omega_{\rm rot}$.
In the present work, we try to reproduce the $B(E2)$ data by
varying the $\gamma$ parameter linearly from
$\gamma=18^\circ$, i.e., $\gamma(\mbox{den})\approx 12^\circ$,
at $\omega_{\rm rot}=0.2$ MeV/$\hbar$ to
$\gamma=30^\circ$, i.e., $\gamma(\mbox{den})\approx 22^\circ$,
at $\omega_{\rm rot}=0.4$ MeV/$\hbar$.
The results are shown as dashed lines in Figs.~\ref{fig:BE2indel}
and~\ref{fig:BE2Rdel}, where rather good agreements for both
the in-band $B(E2)$ and the $B(E2)$ ratio are obtained.
Considering that the various Nilsson Strutinsky calculations~\cite{Bengt},
including ours, predict that the triaxiality parameter does not
change so largely, we need to look for more reliable methods
to determine the mean-field parameters
if this amount of change of the triaxial deformation is really true.

\subsection{Systematic calculations in Lu and Hf nuclei}
\label{sect:systematic}

\begin{figure}[htbp]
\includegraphics[angle=0,width=70mm,keepaspectratio]{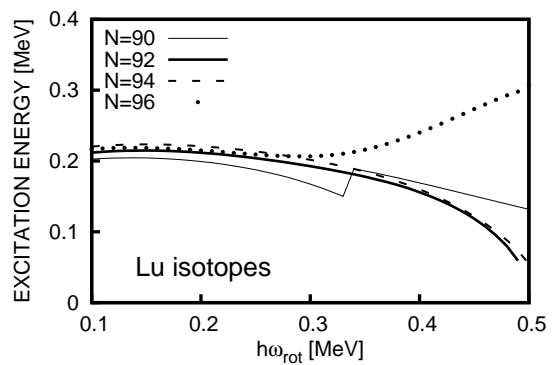}
\caption{
 The eigenenergies of the wobbling-like RPA solutions in Lu isotopes ($Z=71$)
 with $N=$90, 92, 94, and 96
 as functions of the rotational frequency.
 The mean-field parameters are fixed and the same as in Fig.~\ref{fig:crJx}.
 }
\label{fig:RPAomLu}
\end{figure}

\begin{figure}[htbp]
\includegraphics[angle=0,width=70mm,keepaspectratio]{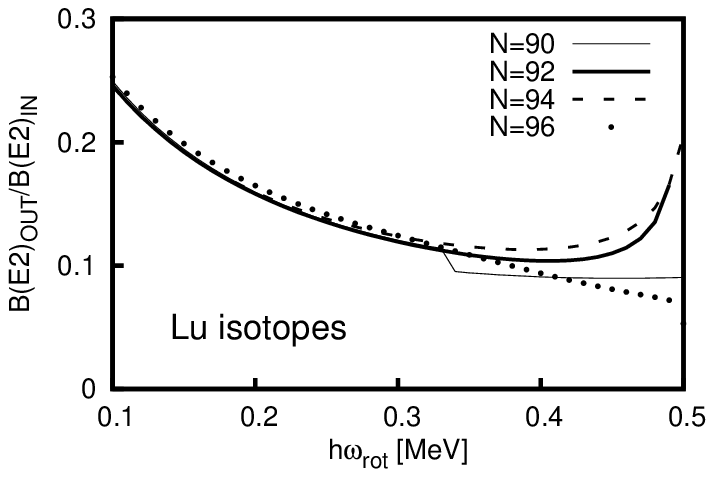}
\caption{
 The out-of-band to in-band $B(E2)$ ratios in Lu isotopes
 with $N=$90, 92, 94, and 96
 as functions of the rotational frequency.
 The mean-field parameters are fixed and the same as in Fig.~\ref{fig:crJx}.
 }
\label{fig:BE2RLu}
\end{figure}

\begin{figure}[htbp]
\includegraphics[angle=0,width=70mm,keepaspectratio]{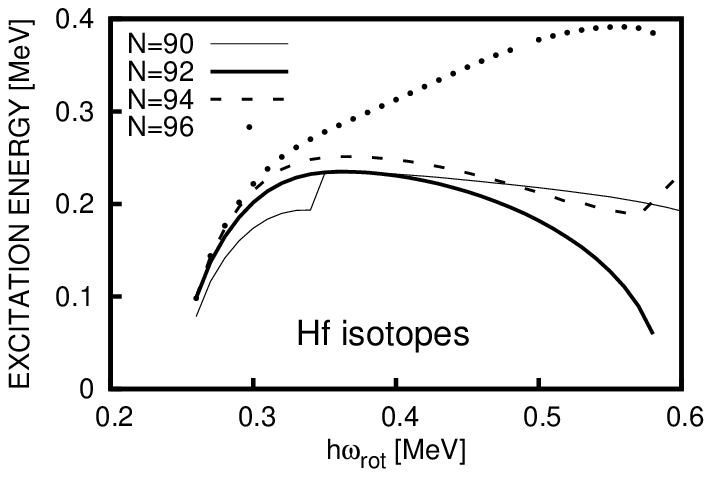}
\caption{
 The eigenenergies of the wobbling-like RPA solutions in Hf isotopes ($Z=72$)
 with $N=$90, 92, 94, and 96
 as functions of the rotational frequency.
 The mean-field parameters are fixed and the same as in Fig.~\ref{fig:crJx}.
 }
\label{fig:RPAomHf}
\end{figure}

\begin{figure}[htbp]
\includegraphics[angle=0,width=70mm,keepaspectratio]{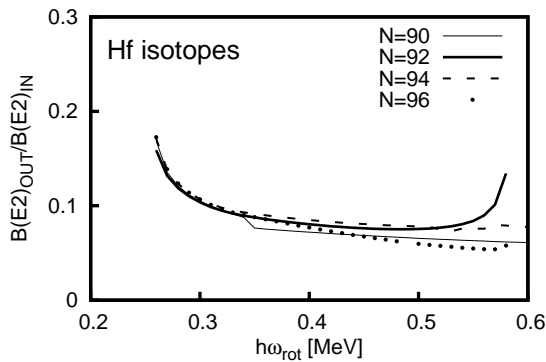}
\caption{
 The out-of-band to in-band $B(E2)$ ratios in Hf isotopes
 with $N=$90, 92, 94, and 96
 as functions of the rotational frequency.
 The mean-field parameters are fixed and the same as in Fig.~\ref{fig:crJx}.
 }
\label{fig:BE2RHf}
\end{figure}

Until now we have considered only the nucleus $^{163}$Lu as a best example.
However, the wobbling excitations have been identified
in several neighboring isotopes, $^{161}$Lu~\cite{wob161Lu},
$^{165}$Lu~\cite{wob165Lu}, and $^{167}$Lu~\cite{wob167Lu}.
Moreover, as it was predicted~\cite{Rag89,Ab90,Bengt},
the many possible candidates of the TSD bands have been
observed in even-even
Hf isotopes~\cite{TSD168Hf,TSD170Hf1,TSD170Hf2,TSD172Hf,TSD174Hf}.
Therefore, it is interesting to know that the existence of
the wobbling excitation is specific for the observed Lu isotopes
or a general phenomenon.
In our microscopic RPA formalism, the wobbling mode is a collective
excitation mode like the low-lying collective vibrations,
the $\beta$- or $\gamma$-vibrations, which are well-known to exist
systematically near the ground states in the wide range of the periodic table.
It can be inspected that the wobbling-like RPA solutions
will be found generally in TSD nuclei.
Thus, we have performed systematic RPA calculations for several
Lu and Hf isotopes with $N=90-96$.
Although the mean-field parameters must be different in each isotope,
we cannot determine them so that we have used for all nuclei
the fixed reference values that are used for $^{163}$Lu
in Sec.~\ref{sect:calRPA}, i.e.,
$(\beta_2,\gamma,\beta_4)=(0.42,18^\circ,0.02)$ and
${\mit\Delta}_\pi={\mit\Delta}_\nu=0.5$ MeV.
The results are shown in the following four figures:
The wobbling excitation energies of the Lu and Hf isotopes are shown
in Figs.~\ref{fig:RPAomLu} and~\ref{fig:RPAomHf}, and
the out-of-band to in-band $B(E2)$ ratios of the Lu and Hf isotopes
in Figs.~\ref{fig:BE2RLu} and~\ref{fig:BE2RHf}, respectively.

As it is clear in these figures, the obtained wobbling excitation
energies are rather similar for the isotopes,
and are about 180 (220) keV in the Lu (Hf) isotopes,
with the exception of $N=96$ isotope, whose excitation energy increases
by about 100$-$200 keV in $\omega_{\rm rot} >$ 0.3 MeV/$\hbar$.
The reason why the excitation energy increases in the $N=96$ isotope 
is that an additional two quasineutron alignment occurs gradually
around that rotational frequency
in this particularly chosen mean-field parameters.
We do not think that it is very meaningful; in fact the observed
wobbling excitation energy in $^{167}$Lu is very similar to that in $^{163}$Lu,
and we have checked that the slight modification of the deformation parameters
changes this trend of increasing energy of the $N=96$ isotopes.
The small kinks of $N=90$ isotopes
at $\omega_{\rm rot} \approx 0.32-0.35$ MeV/$\hbar$
are caused by a sharp crossing of two quasineutrons, whose orbital has
rather small alignment and affects the vacuum very little.
This is also a kind of artifact due to the mean-field parameters
used presently, since the sharp crossing is not observed in experiment
for the $^{161}$Lu isotope.
The $B(E2)$ ratios in the Lu and Hf isotopes are also very similar,
again except for $N=96$ isotopes, whose $B(E2)$ reduce slightly
at higher rotational frequency due to the quasineutron alignment.
In the Lu isotopes, the odd high-$j$ ($N_{\rm osc}=6$) quasiproton
is always occupied.  On the other hand the vacuum states are
zero quasiproton states in the Hf isotopes, and the lowest
two quasiprotons align at $\omega_{\rm rot} \approx 0.23$ MeV/$\hbar$
with relatively large interactions.
As it is discussed in Refs.~\cite{MSMwob1,MSMwob3},
the proton quasiparticle alignment is an important factor
to stabilize the RPA wobbling excitation for the positive $\gamma$ shape,
so that the wobbling-like
solutions appear only after the two quasiproton alignment
in the Hf isotopes.

It must be emphasized that the collective RPA excitation
exists irrespective of the proton number being odd or even
in our microscopic RPA formalism.
Therefore, the wobbling excitations should be identified also
in the even-even Hf nuclei.
It is mysterious for us that there has been no sign of them
in any Hf isotopes:
This may suggest that some important elements may be still missing,
which we do not know yet.

\section{Summary}
\label{sect:sum}

We have investigated the nuclear wobbling motion in the Lu and Hf region
by using the microscopic framework but in the small amplitude approximation,
i.e., the cranked mean-field and the random phase approximation (RPA).
The concept of the wobbling motion is intimately connected to
the rotor model: The body-fixed frame or the principal-axis (PA) frame,
where the degrees of freedom corresponding to the collective rotations
are eliminated, plays a key role.
In the conventional description, i.e., in the uniformly rotating (UR) frame
of the cranking prescription, in which no concept of the body-fixed
frame is introduced, the wobbling excitation is just like an usual collective
vibration; only after transforming into the PA frame it can be
interpreted as a motion of the angular momentum vector wobbling
about the main rotation axis.
Therefore, we reviewed the microscopic RPA framework~\cite{Mar79},
where the PA frame is introduced,
from a slightly general view point~\cite{SM95}:
The original framework by Marshalek~\cite{Mar79} assumes
that the deformation of the mean-field is purely of quadrupole type,
but such a limitation should be lifted for general mean-field potentials.
By introducing the symmetry-preserving residual interaction
associated with a given mean-field potential, it is shown that
the original Marshalek formulation can be recovered with a slight modification.
Furthermore, the strength of this residual interaction is uniquely
determined for a given mean-field so that there is no new parameter
in the calculation of the RPA step, i.e., the theory always has
the consistency between the mean-field and the residual interaction.

As a mean-field, in the present work,
we have used the new Woods-Saxon potential that gives
correct density distributions for both protons and neutrons.
It is believed to give better descriptions than, e.g.,
the Nilsson potential, which was employed in the previous
works~\cite{MSMwob1,MSMwob2,MSMwob3}.
It has been found that the RPA solutions, which can be interpreted
as wobbling excitations, exist systematically in the Lu and Hf region,
if the proper mean-field parameters are adopted, i.e., the pronounced
triaxial deformation or the triaxially superdeformed (TSD) shape.
The general features of the calculated RPA solutions
are rather similar to those obtained in the previous works
employing the Nilsson potential,
except for the $B(M1)$ ratio that is too large,
and the stability against the change of the mean-field parameters
are demonstrated in details.
It should be emphasized that the obtained RPA solutions
satisfy the properties expected from the rotor model;
especially the large out-of-band $B(E2)$ is reproduced
if the same triaxial deformation as the particle-rotor model
analysis~\cite{Ham02,HH03,TST06,TST08} is used.
All these results confirm the wobbling excitations in the TSD bands
from the microscopic view point.

We must admit that the results of our calculations do not
agree very well with the experimental data of the Lu isotopes.
The excitation energies of the wobbling excitation are
underestimated about 150$-$200 keV, which may suggest that
the particle-rotor coupling effect is not fully taken into account.
We obtain the similar excitation energies for wobbling mode
in the even-even Hf nuclei.  Therefore, it is crucial to observe
the wobbling excitations in even-even nuclei to draw a definite
conclusion for whether the effect of the particle-rotor coupling
is essential for the wobbling excitation observed in the odd-$Z$ Lu isotopes.
As for the out-of-band to in-band $B(E2)$ ratio, the calculated
results decrease as functions of the rotational frequency
just like in the rotor model,
if the deformation parameters are kept constants.
The measured $B(E2)$ ratio in $^{163}$Lu is almost constant,
and one has to increase the triaxiality parameter considerably
as a function of the rotational frequency
in order to reproduce this behavior.
However, the Nilsson Strutinsky calculation gives almost constant
triaxial deformation and the obtained triaxiality is too small
to account for the average magnitude of the $B(E2)$ ratio.
The better microscopic description of the wobbling motion
is certainly a challenge to the existing microscopic theory.

\begin{acknowledgments}
We are deeply indebted to Ramon Wyss for providing us
with the new parameter set of the Woods-Saxon potential,
which makes us possible to perform the present work.
\end{acknowledgments}

\end{document}